\documentstyle{article}
\begin{document}
\null
\vspace{1.5cm}
\begin{Large}
{\noindent}{\bf Gravitational Lensing in the Universe}\\
\end{Large}

\bigskip

{\noindent}{\bf XIANG-PING WU}\\

{\noindent}{\bf Beijing Astronomical Observatory, Chinese Academy of
Sciences, Beijing 100080, China}\\

\bigskip

\begin{small}
This work reviews the basic theoretical aspects, the main observational
evidences and the recent applications of gravitational lensing in the
Universe. The article is aimed particularly at providing the readers
who don't work on gravitational lensing a relatively easy introduction
to this active research field in today's astrophysics. \\
\end{small}

\medskip

{\noindent}{\bf Contents}\\

{\noindent}Introduction \\
{\noindent}1 ~~Basic theory\\
	    \null\qquad 1.1 ~~Deflection of light\\
	    \null\qquad 1.2 ~~Lensing geometry and various lenses\\
	    \null\qquad 1.3 ~~Lensing efficiency\\
{\noindent}2 ~~Compact Objects and Microlensing\\
	    \null\qquad 2.1 ~~Microlensing\\
	    \null\qquad 2.2 ~~MACHO searches\\
	    \null\qquad 2.3 ~~Simulations and observations\\
	    \null\qquad 2.4 ~~Inhomogeneous Universe\\
{\noindent}3 ~~Galaxies and Multiple Images\\
	    \null\qquad 3.1 ~~Multiply-imaged Quasars\\
	    \null\qquad 3.2 ~~Multiply-imaged radio sources and radio rings\\
	    \null\qquad 3.3 ~~Galaxies as lenses \\
	    \null\qquad 3.4 ~~Determination of $H_0$\\
	    \null\qquad 3.5 ~~Quasar-galaxy associations\\
{\noindent}4 ~~Clusters of Galaxies and Arclike Images \\
	    \null\qquad 4.1 ~~Giant arcs and arclets \\
	    \null\qquad 4.2 ~~Clusters as lenses \\
	    \null\qquad 4.3 ~~Cluster matter distributions from arcs \\
	    \null\qquad 4.4 ~~Cluster matter distributions from statistical
                              lensing\\
	    \null\qquad 4.5 ~~Cluster matter distributions from weak lensing\\
	    \null\qquad 4.6 ~~Quasar-cluster associations\\
{\noindent}Final remarks \\
{\noindent}Acknowledgements \\
{\noindent}References \\

\bigskip
%@@@@@@@@@@@@@@@@@@@@@@@@@@@@@@@@@@@@@@@@@@@@@@@@@@@@@@@@@@@@@@@@@@@@@@@
\bigskip
{\noindent}{\bf INTRODUCTION}\\

{\noindent}One of the most important tasks for astronomers and physicists
is to study the matter distribution in the Universe.
Based on the assumption of ``light-traces-mass", the map of the Universe
can be directly drawn from the measurements of the apparent positions
of various luminous objects on the sky (two dimensions) and their
distances from the earth (one dimension).
This method has been widely used nowadays and has revealed the existence of
large-scale structures, such as the ``Great Wall", voids, filaments, etc.
However, the dark matter puzzle in today's physics and astrophysics casts
doubt on the hypothesis of using luminous objects as the tracers of the
total matter distribution of the Universe.  Indeed, astronomical observations
can only give rise to the distributions of those celestial objects that have
electromagnetic radiation strong enough to be captured by telescopes,
which might not reflect at all the real matter distribution
in the Universe.

Dynamical methods are traditionally used for the determination of the
masses gravitationally bound in the celestial bodies, which have
successfully led to the discoveries of the excess of dynamical masses
in galaxies and clusters of galaxies as compared with their luminous masses.
Nevertheless, the employment of dynamical analysis
requires that the systems are in the state of  dynamical equilibrium,
while it has remained unclear to date whether the large
gravitationally-dominated
systems like clusters of galaxies have reached the virialized
cosmo-dynamical state, especially at high redshifts.
Furthermore, the mass determination using dynamical
method relies on the detailed knowledge of the celestial bodies:
The rotational velocities and/or velocity dispersions should be
well measured in galaxies and galaxy clusters in order to estimate their
dynamical masses. Yet, this turns out to be quite difficult for the distant
galaxies and galaxy clusters.

A new method of mapping the total matter distribution
(luminous+dark) in the Universe stems from the effect of gravitational
lensing, which has been available only for about 16 years since the
discovery of the first gravitationally lensed double quasar 0957+561A,B
(Walsh, Carswell and Weymann, 1979). The masses derived
from gravitational lensing reflect the total matter contained in
the lensing objects,
independent of whether or not the lensing systems have
reached the virial equilibrium.
Therefore, gravitational lensing provides an independent way to test
the evolution and to set constraints on the
possible form of matter distributions of the lensing objects.
Moreover, studies of gravitational lensing open a possibility of
weighing the unseen matter in the Universe by comparing the
gravitational masses deduced from lensing with the luminous masses
estimated from optical observations.
Furthermore, the effect of gravitational lensing magnifies the apparent
luminosities of background objects, making the intrinsically faint
sources enter into the detection thresholds of telescopes, which
acts in fact like a ``gravitational telescope".

Recall that modern cosmology is based on the so-called
``{\it cosmological principle}", which assumes a spatially
isotropic and homogeneous matter distribution of the Universe.
The isotropy of the Universe has been well demonstrated by the measurement
of the $3K$ microwave background radiation (Smoot et al., 1992),
whilst the homogeneity turns to be
somewhat hard to describe quantitatively. Actually, it is
not so clear on what scales the Universe can be treated as
homogeneous,  though the largest coherent structure seems to have scale of
$\sim100$ Mpc.
Matter condensations occur on scales up to a few tens of megaparsecs:
planets, stars, galaxies, galaxy clusters, superclusters, voids,
Great Walls, etc. All these matter clumps may affect the propagation of light
through the effect of gravitational lensing,  according to
the prediction of the theory of general relativity.  Therefore,
gravitational lensing is a common phenomenon in astronomical
observations and reasonable caution should be exercised in the
identification of various celestial bodies.
For instance, some close double or multiple images may be due to
single sources (e.g. quasars) gravitationally split by the intervening
objects (e.g. galaxies),
and the arclike images could be the result of the gravitationally
distorted background galaxies by the foreground galaxy clusters.

The history of studies of gravitational lensing can
be divided into three periods:  $1704$ -- $1964$, $1964$ -- $1979$ and
$1979$ -- present. Newton (1704) addressed the question nearly
three hundreds years ago if the celestial bodies could bend
light rays. The deflection of light by a spherical body of mass $M$
based on the Newtonian mechanics was computed to be
$\alpha=2GM/c^2\xi$,
assuming an impact distance of $\xi$, i.e., the shortest distance
from $M$ to the light path. $c$ and $G$ are the speed of light and
the gravitational constant, respectively. This formula was found
to underestimate the deflection angle by a factor of $2$
by Einstein in 1915 in terms of his gravitational theory.
In particular, utilizing the result of general relativity to
the Sun predicts a deflection angle of $1.75^{\prime\prime}$  for a light ray
passing near the solar limb.  This prediction was very soon confirmed
during the solar eclipse in 1919 by a team led by Eddington
(Dyson, Eddington and Davidson, 1920). Although
some progress had been made since then on the theoretical aspects
of light bending, not much interest had been really drawn on
this field until 1963--1964 when the geometry of lensing was studied
independently by Klimov (1963),
Liebes (1964) and Refsdal (1964a).  Their work can
be considered pioneering in the sense that they set up the
foundation of the modern theoretical research on gravitational
lensing. During the period of 1964 -- 1979 some important progress
was made in the computations of deflection angle
(e.g., Bourassa and Kantowski, 1975; 1976) and the light propagation in
the model of an inhomogeneous Universe
(Press and Gunn, 1973; Dyer and Roeder, 1972; 1973; 1974).
These achievements have played an important role in the lensing
studies after the detection of the first multiple images of
quasar 0957+561A,B by Walsh, Carswell and Weymann in 1979. In the decade
following this landmark, the detections of
new lensing phenomena have increased dramatically,
including multiply-imaged quasars, giant luminous arcs and arclets,
radio Einstein rings, microlensing events,
whilst the theoretical investigation
of gravitational lensing has concerned many interesting subjects of modern
cosmic physics, such as determinations of total masses of the lensing
systems,  determinations of $H_0$ and $\Omega_0$, explanation of the
associations of background sources with foreground objects, searches for
dark matter candidates, etc. Today, gravitational lensing
has become one of the most active fields in cosmological research.

It is surely impossible to cover all the topics of lensing research in
this review as it has expanded very rapidly in the past years.
The purpose of this work is to concentrate on the
basic theories and the new progress of gravitational
lensing as well as their applications in astrophysics.  Alternatively,
this review will not trace the history of development of
gravitational lensing but follow the scales of lenses from compact
objects to large-scale inhomogeneities.

\bigskip
%@@@@@@@@@@@@@@@@@@@@@@@@@@@@@@@@@@@@@@@@@@@@@@@@@@@@@@@@@@@@@@@@@@@@@@@
\bigskip
\section{BASIC THEORY}

\subsection{Deflection of light}
Gravitational lensing is based on the theory of general relativity, which
predicts that light rays would be bent when they pass near a massive
body. For a pointlike mass $M$ the deflection angle of light rays is
%1
\begin{equation}
\alpha=\frac{4GM}{c^2\xi},
\end{equation}
where $\xi$ represents the impact distance of light rays. This deflection
angle is twice as large as the value predicted from Newtonian mechanics.

For an extended mass distribution, the deflection of light passing through
the mass system cannot be simply obtained in the frame of general
relativity.  Only for some special matter distributions can the metric
and the solution
to the photon geodesic equations, hence the angle of light bending,
be found exactly. For instance, in a spherical uniform matter
distribution the deflection of light, to  first order in the Newtonian
gravitational potential, is  (Wu, 1989a)
%2
\begin{equation}
\alpha\approx\frac{4GM}{c^2\xi}
\left[1-\left(1-\frac{\xi^2}{R^2}\right)^{3/2}\right].
\end{equation}
Here $M$ and $R$ are the total mass and the radius of the massive sphere,
respectively. Another matter profile that has been commonly used for galaxies
and clusters of galaxies is the singular isothermal sphere model
having mass density
%3
\begin{equation}
\rho(r)=\frac{\sigma_{v}^2}{2\pi G r^2},
\end{equation}
where $\sigma_{v}$ measures the line-of-sight velocity dispersion. The
deflection angle of light rays can be found to  first order in
$(\sigma_{v}/c)^2$ to be (Wu, 1989a)
%4
\begin{equation}
\alpha=4\pi\frac{\sigma_{v}^2}{c^2},
\end{equation}
i.e., a constant deflection independent of the impact distance.

For photons traveling in the gravitational field of
an irregular matter distribution, the deflection angle of light rays
is often computed in a linearized Einstein approximation
%5
\begin{equation}
\mbox{\boldmath $\alpha$}=
\frac{2}{c^2}\int_{-\infty}^{\infty}\bigtriangledown \phi dt,
\end{equation}
in which $\phi$ represents the Newtonian gravitational potential of
the matter distribution. An equivalent but simple formula is obtained by
dividing the system into a number of pointlike masses ($M_i$) and then
summing up the contributions from each small mass piece
%6
\begin{equation}
\mbox{\boldmath $\alpha$}=\displaystyle \sum_{i}\frac{4GM_i}{c^2}
\frac{{\bf r}-{\bf r_i}}{|{\bf r}-{\bf r_i}|^2},
\end{equation}
where ${\bf r}-{\bf r_i}$ is the impact vector of light
rays from the mass unit $M_i$. The integral form is
%7
\begin{equation}
\mbox{\boldmath $\alpha$}=\frac{4G}{c^2}\int
\frac{ \mbox{\boldmath $\xi$} - \mbox{\boldmath $\xi^{\prime}$} }
     {|\mbox{\boldmath $\xi$} - \mbox{\boldmath $\xi^{\prime}$} |^2}
    \Sigma(\mbox{\boldmath $\xi^{\prime}$})
    d^2\mbox{\boldmath $\xi^{\prime}$}.
\end{equation}
Here $\Sigma(\mbox{\boldmath $\xi^{\prime}$})$
is the surface mass density at position $\mbox{\boldmath $\xi^{\prime}$}$
obtained by projecting all the mass along the line of sight onto the
``lens plane".  In particular,  for a spherical matter distribution $\rho(r)$
the above equation can be simplified to be
%8
\begin{equation}
\alpha=\frac{4Gm(\xi)}{c^2\xi},
\end{equation}
where $m(\xi)$ is the total projected mass along the light-of-sight
enclosed within the impact distance $\xi$. If $\rho(r)$ is confined
within the radius $R$, then
%9
\begin{equation}
\begin{array}{ll}
m(\xi)=M-4\pi\int_{\xi}^R\sqrt{r^2-\xi^2}r\rho(r) dr, \;\;\;& \xi<R;\\
m(\xi)=M, &   \xi\geq R,
\end{array}
\end{equation}
in which $M$ is the total mass of the spherical system.
Therefore, if the deflector has  spherical symmetry, the light bending
can be obtained simply  by replacing the pointlike mass $M$ of eq.(1)
by the projected mass $m(\xi)$.  For example, replacing the mass
density in eq.(9)  by a constant and inserting $m(\xi)$ for
$\xi<R$  into eq.(8) recover the result of eq.(2).
It should be pointed out that eq.(9) holds true even if the deflector
has no boundary or infinite mass, in which
%10
\begin{equation}
m(\xi)=M(\xi)+4\pi\int_{\xi}^{\infty}(r-\sqrt{r^2-\xi^2})r\rho(r) dr.
\end{equation}
So, the deflection angle [eq.(4)] of a singular isothermal sphere
can be easily obtained by inserting the density profile eq.(3)
into eqs.(8) and (10).\\

%----------------------------------------------------------------------
\subsection{Lensing geometry and various lenses}

Suppose that a light-ray from a distant source at redshift $z_s$ passes through
or near a massive system at redshift $z_d$ ($z_d<z_s$) and then reaches  the
observer at redshift $z=0$ (Figure 1).  If there were no intervening
massive system, the light from the source would have
arrived at the observer along a straight line.
%@@@@@@@@@@@@@@@@@@@@@@@@@@@@@@@@@@@@@@@@@@@@@@@@@@@@@@@@@@@@@@@@@@@@@@
%                          FIGURE 1
\begin{figure}
 \vspace{1cm}
 \caption{Scheme of basic geometry of gravitational lensing.
	  The true position of the distant source is $\beta$ with
	  an angular diameter distance $D_s$ from the observer.
	  The light rays with an impact distance $\theta$
	  would be deflected by an angle  $\alpha$ by the
	  gravitational field of an intervening massive object, resulting
          in multiple and distorted images.}
\end{figure}
%@@@@@@@@@@@@@@@@@@@@@@@@@@@@@@@@@@@@@@@@@@@@@@@@@@@@@@@@@@@@@@@@@@@@@@
The gravitational field of the massive system now bends the
light ray, causing the direction of light to be changed by an angle of
$\mbox{\boldmath $\alpha$}$. If we use $\mbox{\boldmath $\eta$}$
($\mbox{\boldmath $\beta$}$ in angle) to denote the
true position of the source or the alignment parameter in the source
plane and $\mbox{\boldmath $\xi$}$ ($\mbox{\boldmath $\theta$}$ in angle),
the observed positions of the images
in the lens plane, we have the following geometrical relation, namely,
the ``lensing equation":
%11
\begin{equation}
\frac{D_s}{D_d}\mbox{\boldmath $\xi$}-\mbox{\boldmath $\eta$}\;
=\;D_{ds} \mbox{\boldmath $\alpha$},
\end{equation}
or
%12
\begin{equation}
\mbox{\boldmath $\theta$}-\mbox{\boldmath $\beta$}\;
=\;\frac{D_{ds}}{D_s}\mbox{\boldmath $\alpha$}\;
\end{equation}
where $D_d$, $D_s$ and $D_{ds}$ are the angular diameter distances
to the deflector, to the source and from the deflector to the
source, respectively.  As seen from eq.(7), the deflection angle
$\mbox{\boldmath $\alpha$}$ can be written as the gradient of a
two-dimensional potential $\psi$
%13
\begin{equation}
\frac{D_{ds}}{D_s}\mbox{\boldmath $\alpha$}=\nabla \psi
\end{equation}
and
%14
\begin{equation}
\psi(\mbox{\boldmath $\theta$})=\int\;
\frac{\Sigma({\mbox{\boldmath $\theta^{\prime}$}})}{\pi\Sigma_c}
\ln |\mbox{\boldmath $\theta$}-\mbox{\boldmath $\theta^{\prime}$}|
\;d^2\mbox{\boldmath $\theta^{\prime}$},
\end{equation}
in which the quantity $\Sigma_c$ is called the critical
surface mass density
%15
\begin{equation}
\Sigma_c=\frac{c^2}{4\pi G}\frac{D_s}{D_dD_{ds}},
\end{equation}
and the physical surface mass density term $\Sigma$ should satisfy
the two dimensional Poisson's equation (Blandford and Narayan, 1986)
%16
\begin{equation}
\bigtriangledown^2\psi=\frac{2\Sigma}{\Sigma_c}\equiv2\kappa.\\
\end{equation}

For a given source position $\mbox{\boldmath $\beta$}$,
the lensing equation may have several solutions for the
image position $\mbox{\boldmath $\theta$}$.
As a consequence, we can observe multiple images of a single source
on the sky.  Furthermore, the lensing equation (12) describes the
distortion of the surface brightness of background sources according to
the mapping from source plane to lens  plane:
%17
\begin{equation}
I^{\prime}(\mbox{\boldmath $\theta$})=I(\mbox{\boldmath $\beta$})
=I(\mbox{\boldmath $\theta$}-\nabla \psi),
\end{equation}
where $I^{\prime}$  and $I$ are, respectively,
the observed and intrinsic surface
brightness patterns.  The magnification factor $\mu$ describes
the change of apparent luminosity of the source, which is characterized by
the Jacobian for the mapping $\mbox{\boldmath $\beta$}$ $\longrightarrow$
\mbox{\boldmath $\theta$} (Schneider, Ehlers and Falco,  1992):
%18
\begin{equation}
\mu\equiv \left| {\rm det}\frac{\partial \mbox{\boldmath $\beta$}}
{\partial \mbox{\boldmath $\theta$}}\right|^{-1}
=\left[ \left(1-\kappa\right)^2-\gamma^2 \right]^{-1},
\end{equation}
here
%19
\begin{equation}
\begin{array}{lll}
\kappa & = & \frac{1}{2}(\psi_{11}+\psi_{22});\\
 &   & \\
\gamma & = & \sqrt{\gamma_1^2+\gamma_2^2};\\
 &   & \\
\gamma_1 & = & \frac{1}{2}(\psi_{11}-\psi_{22});\\
 &   & \\
\gamma_2 & = & \psi_{12}=\psi_{21},\\
\end{array}
\end{equation}
in which $\psi_{ij}\equiv\partial^2\psi/\partial\theta_i\partial\theta_j$.
When the line of sight completely misses the deflector,
$\Sigma(\mbox{\boldmath $\theta$})=0$ and the $\kappa$ term vanishes
in terms of eq.(16).
So, $\kappa$ represents the amplitude of the convergence due to the matter
within the light-ray (also referred to as Ricci focusing), while
the $\gamma$ term is the amplitude of the shear due to the matter outside
the beam (also referred to as Weyl focusing).  The latter can be easily
calculated in the case of spherical deflector, which reads
%20
\begin{equation}
\gamma=\frac{4G[m(\theta)-\overline{m}(\theta)]}
{c^2\theta^2}\frac{D_{ds}}{D_dD_s},
\end{equation}
and $\overline{m}(\theta)=\pi \theta^2D_d^2\Sigma(\theta)$
is the mass of the cylinder of radius
$\theta$ with  a uniform surface mass density equal to $\Sigma(\theta)$.
When a beam of light-rays pass through the center of deflector,
$\overline{m}(0)=m(0)=0$ and the $\gamma$ term then becomes zero.

\underline{(1)Pointlike mass as deflector.}
This model can be considered to be a good approximation for many celestial
bodies like ``Jupiters", stars, black holes, and even galaxies, when the
light rays from background sources pass outside the deflectors.
Solving the lensing equation, using the deflection of eq.(1), yields
%21
\begin{equation}
\theta_{\pm}=\frac{1}{2}\left(\beta\pm\sqrt{\beta^2+4\theta_{E}^2}\right),
\end{equation}
where
%22
\begin{equation}
\theta_{E}=\left(\frac{4GM}{c^2}\frac{D_{ds}}{D_d D_s}\right)^{1/2}
\end{equation}
is the critical radius (or $a_{E}=\theta_{E}D_d$ in linear size)
corresponding to a ring-like image of a background source when $\beta=0$,
which is often called the ``Einstein ring", named after the pioneering work
by Einstein (1936).  $\theta_+$ and $\theta_-$ describe the positions
of the two images produced by a point mass $M$. Their magnifications are
%23
\begin{equation}
{\mu}_{\pm}=\left|\frac{1}{1-(\theta_{E}/\theta_{\pm})^4}\right|.
\end{equation}
If the separation between the two images is too small to be resolved
by modern telescopes, as for the microlensing events (section 2),
the total magnification is often used for their combined effect
%24
\begin{equation}
{\mu}={\mu}_++{\mu}_-=\frac{u^2+2}{u\sqrt{u^2+4}}.
\end{equation}
Here $u$ is defined as $u=\beta/\theta_{E}$. $u=1$ or $\beta=\theta_{E}$ is
often taken to be a typical case that characterizes the efficiency of the
lens, which corresponds to $({\mu}_+,{\mu}_-)=(1.17,0.17)$ and
${\mu}=1.34$ or $\Delta m=0.32$ in apparent magnitude.
%The variation of ${\mu}$ with the dimensionless impact parameter
%$u$ is shown in Figure 2.
Pointlike masses as lenses play
an important role in the study of microlensing (section 2.1)
and in searches for dark matter candidates in the Galactic
halo using microlensing effect (section 2.2).

The image configurations by a pointlike mass is well illustrated
in Figure 2 by changing the relative positions between the source
%@@@@@@@@@@@@@@@@@@@@@@@@@@@@@@@@@@@@@@@@@@@@@@@@@@@@@@@@@@@@@@@@@@@@@@
%                          FIGURE 2
\begin{figure}
 \vspace{1cm}
 \caption{Image configurations by a black hole
          (Figure courtesy of C. Zahn and H. Ruder)}
\end{figure}
%@@@@@@@@@@@@@@@@@@@@@@@@@@@@@@@@@@@@@@@@@@@@@@@@@@@@@@@@@@@@@@@@@@@@@@
which is chosen to be the Einstein portrait
and the lens which is assumed to be a massive black hole. When
the source approaches the lens, the two images are elongated,
showing two arclike structures. Finally, two arc images merge into
an ``Einstein ring" when the lens is completely coincident with the source.\\

\underline{(2)A singular isothermal sphere.}  The studies of the
flat rotation curves of galaxies and the galaxy/gas distributions in clusters
of galaxies suggest that the total matter profiles in these systems follow
very well the singular isothermal sphere (SIS) model [see eq.(3)].
The constant deflection
of $\alpha=4\pi(\sigma_{v}/c)^2$  in the lensing equation gives
%25
\begin{equation}
\theta_{\pm}=\theta_{E}\pm\beta
\end{equation}
for $\beta<\theta_E$, where the critical radius is
%26
\begin{equation}
\theta_{E}=4\pi\frac{\sigma_{v}^2}{c^2}\frac{D_{ds}}{D_s}.
\end{equation}
In particular, the image separation is just the diameter
of the Einstein ring: $\Delta \theta=\theta_++\theta_-=2\theta_{E}$.
Note that if the alignment
parameter $\beta$ is larger than the Einstein ring, only one image appears
instead of two. The magnifications of the images
are simply
%27
\begin{equation}
{\mu}_{\pm}=\left|1\mp\frac{\theta_{E}}{\theta_{\pm}}\right|^{-1}
=\left|1\pm 4\pi\frac{\sigma_{v}^2}{c^2}\frac{D_{ds}}{\beta D_s}\right|.
\end{equation}

\underline{(3)A softened singular isothermal sphere.}~
SIS, though simple, is an unphysical model because the mass density
reaches infinity at the center. Instead, a SIS with a finite core radius of
$r_c$ (or $\theta_c$ in angle) (ISC) seems to be more reasonable
for the matter distributions of galaxies and galaxy clusters
(Hinshaw and Krauss, 1987):
%28
\begin{equation}
\rho(r)=\frac{\sigma_{v}^2}{2\pi G}\frac{1}{r^2+r_c^2}.
\end{equation}
This density profile reduces to SIS when $r_c=0$ or $r\gg r_c$. The surface
mass density and the total projected mass within $\xi$ are
%29
\begin{equation}
\begin{array}{lll}
\Sigma(\xi)& = &\frac{\sigma_{v}^2}{2G}\frac{1}{\sqrt{\xi^2+r_c^2}};\\
  &  &  \\
m(\xi)& =& \frac{\pi \sigma_{v}^2}{G}\left(\sqrt{\xi^2+r_c^2}-r_c\right),
\end{array}
\end{equation}
respectively. The deflection of light can be directly obtained from eq.(9) and
then, the lensing equation reads
%30
\begin{equation}
\theta_0=\beta_0+D\frac{\sqrt{1+\theta_0^2}-1}{\theta_0},
\end{equation}
in which $\theta_0$ and $\beta_0$ are in unit of $\theta_c$.
The lensing parameter,  defined as $D\equiv (4\pi\sigma_v^2/
c^2)(D_dD_{ds}/r_cD_s)$,
determines the number of the solutions (Figure 3) (Wu, 1989b).
In the case of $D\leq2$, ISC always produces a single image,
while  it may result in three images for $D>2$ if $\beta_0$ is
sufficiently small.
The intersections of the lines $\beta_0=$constant
with the curves $\beta_0=\theta_0-D(\sqrt{1+\theta_0^2}-1)/\theta_0$
give the solutions to the lensing equation.
%@@@@@@@@@@@@@@@@@@@@@@@@@@@@@@@@@@@@@@@@@@@@@@@@@@@@@@@@@@@@@@@@@@@@@@
%                          FIGURE 3
\begin{figure}
 \vspace{1cm}
 \caption{ Solution to lensing equation by ISC. The solid curves
	  represent the lensing equation
	  $\beta_0 = \theta_0 - D[(1+\theta_0^2)^{1/2}-1]/\theta_0$.
	  The intersecting points with the line $\beta_0=$constant give
	  the number and positions of the lensed images. }
\end{figure}
%@@@@@@@@@@@@@@@@@@@@@@@@@@@@@@@@@@@@@@@@@@@@@@@@@@@@@@@@@@@@@@@@@@@@@@
The magnification is found to be
%31
\begin{equation}
{\mu}=\left|\left(1-D\frac{\sqrt{1+\theta_0^2}-1}{\theta_0^2}\right)
\left(1+D\frac{\sqrt{1+\theta_0^2}-1}{\theta_0^2}-
D\frac{1}{\sqrt{1+\theta_0^2}}\right)
\right|^{-1}
\end{equation}

\underline{(4)Other spherical models.} Two other models which are also
frequently adopted for the matter distributions in
galaxies and in galaxy clusters are the King model
(or the modified Hubble model)
and the de Vaucouleur model (or the $r^{1/4}$ law).
%********************** TABLE 1 **********************************
   \begin{table}
      \caption{Gravitational Lensing Models}
         \label{ }
      \vspace{1cm}
   \end{table}
%********************** TABLE 1 **********************************
Their lensing properties are very similar to those of ISC and are summarized
in Table 1. \\

\underline{(5)Asymmetric lenses.}~  In principle, the lensing geometry can be
established for any kind of geometrically-thin matter inhomogeneity in
the Universe through
eqs.(7) and (12). Among them, the properties of  elliptical lenses have
been thoroughly studied (e.g., Bourassa and Kantowski, 1975; 1976;
Blandford and Kochanek, 1987; Narasimha, Subramanian and Chitre, 1987;
Schramm, 1990; Wallington and Narayan, 1993; Kassiola and Kovner, 1993, etc.).
A simple elliptical lens assumes the following
``non-singular pseudo-elliptical isothermal potential"
%32
\begin{equation}
\psi(\theta_1,\theta_1)=4\pi\frac{\sigma_v^2}{c^2}\frac{D_dD_{ds}}{D_s}
\sqrt{\theta_c^2+(1-\epsilon)\theta_1^2+(1+\epsilon)\theta_2^2},
\end{equation}
where $\epsilon$ is the ellipticity and $\theta_c$ is the
core radius of the lens.
Solving the lens equation
%33
\begin{equation}
\mbox{\boldmath $\beta$}\;=\;
\mbox{\boldmath $\theta$}\;-\;\bigtriangledown \psi,
\end{equation}
one has (Blandford and Kochanek, 1987; Wallington and Narayan, 1993)
%34
\begin{equation}
\begin{array}{lll}
\beta_1 & = & \theta_1-\theta_E\frac{(1-\epsilon)\theta_1}
{\sqrt{\theta_c^2+(1-\epsilon)\theta_1^2+(1+\epsilon)\theta_2^2}},\\
\beta_2 & = & \theta_2-\theta_E\frac{(1-\epsilon)\theta_2}
{\sqrt{\theta_c^2+(1-\epsilon)\theta_1^2+(1+\epsilon)\theta_2^2}},
\end{array}
\end{equation}
where  $\theta_E$ is the Einstein radius in
SIS [eq.(26)]. Magnifications of the images can be calculated using eq.(18)
%35
\begin{equation}
{\mu}= \left|\frac{\partial \beta_1}{\partial \theta_1}
             \frac{\partial \beta_2}{\partial \theta_2}\; - \;
	     \frac{\partial \beta_1}{\partial \theta_2}
             \frac{\partial \beta_2}{\partial \theta_1}
       \right|^{-1}.
\end{equation}
Inverting eq.(34) gives the image positions in the lens plane
for  pointlike sources. An extended source can be considered
as a set of pointlike sources and its image configuration produced by an
elliptical lens can be drawn through the above equation for each
source element. In principle, the images with arbitrary
%@@@@@@@@@@@@@@@@@@@@@@@@@@@@@@@@@@@@@@@@@@@@@@@@@@@@@@@@@@@@@@@@@@@@@@
%                          FIGURE 4
\begin{figure}
 \vspace{1cm}
 \caption{Image configurations by an elliptical lens. The source planes are on
          the left and the corresponding images are on the right.
          The solid lines are the caustics and the dashed lines are the
          corresponding critical lines. (from Blandford and Narayan, 1992)}
\end{figure}
%@@@@@@@@@@@@@@@@@@@@@@@@@@@@@@@@@@@@@@@@@@@@@@@@@@@@@@@@@@@@@@@@@@@@@@
shapes can be produced  by an elliptical lens as long as
the position and the shape of the extended source are properly chosen.
Some image configurations produced in this
procedure for a circular source are shown in Figure 4.\\

\underline{(6)Cosmic strings as lenses.}~  Cosmic strings are
topological defects that are believed to be created in the very early Universe
and could act as  the seeds of formations of galaxies,  galaxy clusters
and even large-scale structures (Vilenkin, 1981), although there has been
no  observational evidence to date for the existence of such cosmic
strings in the Universe. Cosmic strings, if real, may essentially exist in
two forms: straight and loop strings. Thus, the long-lived cosmic strings
are able to cause gravitational lensing effect on background sources.

The exterior gravitational field of a straight string is ($G/c^2=1$)
(Gott, 1985)
%36
\begin{equation}
ds^2=-dt^2+dr^2+(1-4\mu_s)^2r^2d\phi^2+dz^2.
\end{equation}
Thus, a cosmic string is completely described by its linear mass
density $\mu_s$ ($\sim10^{-6}$). Consider that a light ray propagates
in the plane perpendicular to the string (Figure 5), one can easily
solve the null geodesic equation and find the light deflection in
$\phi$ direction to be
%@@@@@@@@@@@@@@@@@@@@@@@@@@@@@@@@@@@@@@@@@@@@@@@@@@@@@@@@@@@@@@@@@@@@@@
%                          FIGURE 5
\begin{figure}
 \vspace{1cm}
 \caption{Geometry of a straight cosmic string}
\end{figure}
%@@@@@@@@@@@@@@@@@@@@@@@@@@@@@@@@@@@@@@@@@@@@@@@@@@@@@@@@@@@@@@@@@@@@@@
%37
\begin{equation}
\alpha\approx 4\pi \mu_s\;=\;2.6^{\prime\prime}(\mu_s/10^{-6}).
\end{equation}
The light rays  that propagate along $r$ and $z$ directions
are unaffected by the gravitational field of a straight string,
according to the metric of eq.(36).  Two images of a background
source with the separation of  $8\pi\mu_s$ but with
the same apparent luminosity appear around the string. Note that
eq.(36) adopts a cylindrical coordinate rather than the spherical one
as in SIS although both lenses produce the impact parameter-independent
deflections. So, the double images of a background source by a straight
string can be identified relatively easily from their equal luminosities.

The metric of a long-lived loop string is unknown today.
Its lensing properties can be studied only in the linearized
gravitational approximation. For the simple case of
a ``face-on" loop string with radius of $a$ in the sky, the light
from the source behind the string is bent by (Wu, 1989c)
%38
\begin{equation}
\alpha=\left\{
\begin{array}{ll}
0, & \;\;\;\;\xi<a;\\
8 \pi \mu_s a/\xi, & \;\;\;\; \xi>a,
\end{array} \right.
\end{equation}
i.e., the light rays which pass through the loop remain unaffected
while the light rays outside the loop behave in the same way as those by
a point mass of $M=2\pi\mu_s a$ at the center. Therefore, if the alignment
parameter $\eta$ of a background source is smaller than the loop radius $a$,
we would always expect to see the original source through the loop.
In particular, when $\eta$ satisfies
$a-8\pi\mu_s(D_dD_{ds}/D_s)<\eta<a$,  the arclike image of the lensed
source appears outside the loop.
Figure 6 shows such an example, in which the
%@@@@@@@@@@@@@@@@@@@@@@@@@@@@@@@@@@@@@@@@@@@@@@@@@@@@@@@@@@@@@@@@@@@@@@
%                          FIGURE 6
\begin{figure}
 \vspace{1cm}
 \caption{Loop string and arclike image}
\end{figure}
%@@@@@@@@@@@@@@@@@@@@@@@@@@@@@@@@@@@@@@@@@@@@@@@@@@@@@@@@@@@@@@@@@@@@@@
source and loop string are $10$ kpc and $100$ kpc
in radii and located at $z_s=1$ and $z_d=0.25$, respectively.
The equivalent mass of the string is $\sim10^{14}M_{\odot}$.  Therefore,
a single giant arclike image, instead of multiple arclike images in the
case of massive spherical deflectors,  can be produced
by the loop string.\\

%---------------------------------------------------------------------
\subsection{Lensing efficiency}

In the above discussion, it appears that
lensing magnification of the apparent luminosities of background sources
may tend towards infinity in some cases. For
example, the impact parameters that satisfy the condition
$(1-\kappa)^2-\gamma^2=0$ in eq.(18) correspond to images with
an infinitely large magnification, whilst the same situation occurs
for a pointlike lens when the alignment parameter is $\eta=0$.
These apparent unphysical results arise from the hypothesis that
the background source is pointlike.

In general, the total magnification of an extended source with surface
bightness $I(\eta_1,\eta_2)$ can be obtained by summing up the
contribution ${\mu}(\eta_1,\eta_2)$ of each source element
$I(\eta_1,\eta_2)d\eta_1d\eta_2$ (Bontz, 1979):
%39
\begin{equation}
{\mu}=\frac{\int\int {\mu}(\eta_1,\eta_2)I(\eta_1,\eta_2)
       d\eta_1d\eta_2}{\int\int I(\eta_1,\eta_2)d\eta_1d\eta_2}.
\end{equation}
For simplicity, we consider a circular disk source with uniform
surface brightness and  radius $R_s$ instead of the point source
approximation, and we further assume a  spherical
matter distribution for the deflector. In this situation,  the maximum
magnification ${\mu}_{max}$ corresponds to the case where the centers of
the source and of the lens and the observer lie perfectly on a straight
line, i.e, the Einstein ring shows up. So, eq.(39) reduces to the
ratio of the area of the Einstein ring to the luminous area of the
original source.
This ratio or maximum magnification is not infinite any longer. Actually,
${\mu}_{max}$ measures the efficiencies of gravitational lensing by
various lenses for the same source. The larger the ${\mu}_{max}$ is,
the stronger the lens would be. For a pointlike lens $M$,
%40
\begin{equation}
{\mu}_{max}=\sqrt{1+\frac{16GM}{c^2}\frac{D_sD_{ds}}{R_s^2D_d}};
\end{equation}
and for SIS,
%41
\begin{equation}
{\mu}_{max}=\left\{
\begin{array}{ll}
16\pi\frac{\sigma_v^2}{c^2}\frac{D_{ds}}{R_s} \;\;\;& \;\;
                                            \theta_s<\theta_{E}  \\
1+8\pi\frac{\sigma_v^2}{c^2}\frac{D_{ds}}{R_s}\;\;\;& \;\;
                                            \theta_s>\theta_{E},
\end{array}\right.
\end{equation}
where $\theta_s=R_s/D_s$ is the angular radius of the source.

To show how efficiently the various lenses act on the background source,
the maximum magnification of a background circular disk source
is illustrated in
Figure 7 for three typical lenses in the Universe: stars, galaxies
and clusters of galaxies (Wu, 1992a). The star is modeled by a point mass
with a solar mass $M_{\odot}$, while SIS is adopted for galaxies and
galaxy clusters whose velocity dispersions are taken to be 200 km/s and
1500 km/s, respectively.
%@@@@@@@@@@@@@@@@@@@@@@@@@@@@@@@@@@@@@@@@@@@@@@@@@@@@@@@@@@@@@@@@@@@@@@
%                          FIGURE 7
\begin{figure}
 \vspace{1cm}
 \caption{Maximum magnification and source radius. The background source
	is assumed to be a luminous circular  disk
	with radius $R_s$ at $z_s=2$
	while the lens is modeled by a point mass for star [eq.(40)]
	and SIS for galaxy and cluster of galaxy [eq.(41)] at $z_d=0.5$. }
\end{figure}
%@@@@@@@@@@@@@@@@@@@@@@@@@@@@@@@@@@@@@@@@@@@@@@@@@@@@@@@@@@@@@@@@@@@@@@
It is concluded from Figure 7: (1)Compact objects like stars as lenses
are capable of producing significant lensing effect on  sources with
sizes smaller than $\sim0.01$ pc. Therefore, they can affect AGNs
($\sim10^{-3}$ pc), quasars and normal stars; (2)Lenses on scale of
galaxies can magnify any sources with size smaller than galaxies
themselves; (3)Finally, clusters of galaxies are efficient lenses for
nearly all luminous objects including galaxies of sizes of $\sim10$ kpc.\\

\bigskip
%@@@@@@@@@@@@@@@@@@@@@@@@@@@@@@@@@@@@@@@@@@@@@@@@@@@@@@@@@@@@@@@@@@@@@@@
\bigskip
\section{COMPACT OBJECTS AND MICRO-LENSING}

%----------------------------------------------------------------------
\subsection{Microlensing}

For a pointlike lens $M$ and a background source, both at their typical
cosmological distances of $z_d=0.5$ and $z_s=1$, the angle subtended
by the Einstein radius is [eq.(22)]
%42
\begin{equation}
\theta_E\;\approx\; 1.4\times10^{-6}\;
\left(\frac{M}{M_{\odot}}\right)^{1/2} \;{\rm arcseconds}.
\end{equation}
Thus, a compact star-like lens at cosmological distance gives rise to
an image splitting of a background source of the order of
``microarcseconds". Hence, the terms ``microlens" and ``microlensing"  are
used for the compact object and its lensing phenomena so as to distinguish
it from the ``macrolens" like a galaxy which results in the image
separation of the order of $\sim1$ arcsecond according to eq.(42).
Nevertheless, it appears to be hopeless to resolve the images microlensed
by star-like compact objects in the Universe even with the present
advanced telescopes.

Because of the too small separation between the multiply microlensed images,
the combined magnification of the images turns
to be the unique feature that arises from microlensing. Unfortunately,
this feature cannot be straightforwardly used to detect microlensing events
since it is impossible to separate the magnification effect from the intrinsic
luminosity of the source. However, if the source and/or the lens
have a transverse motion with respect to the line of sight,  the lensing
magnification would vary with time, leading to the variation of apparent
luminosity of the source. This effect was firstly predicted by Chang and
Refsdal in 1979. Since then, a great number of papers have appeared, trying to
apply this property for the explanation of QSO variabilities and
the searches of dark matter candidates of the Galactic halo (e.g.,
Young, 1981; Canizares, 1982;  Ostriker and Vietri, 1985; Nottale, 1986;
Paczy\'nski, 1986; Schramm et al., 1994; etc.).
The timescale of luminosity variability
of a source by a microlens $M$ can be estimated using the time of the source
crossing the Einstein radius $a_E$ with a relative velocity $v$:
$T= 2a_E/v$. (1)For a local Galactic lens, e.g. an object in the
Galactic halo ($D_d\sim10$ kpc) acting as lens and a star of the Large
Magellanic Cloud (LMC) ($D_s=50$ kpc) being the target source:
%43
\begin{equation}
T\approx 0.2 {\rm yr}\;\left(\frac{M}{M_{\odot}}\right)^{1/2}
         \left(\frac{v}{200\;{\rm km\;s^{-1}}}\right)^{-1};
\end{equation}
and (2)for a lens at cosmological distance ($z_d=0.5$ and $z_s=1$)
%44
\begin{equation}
T\approx 10 {\rm yr}\;\left(\frac{M}{M_{\odot}}\right)^{1/2}
         \left(\frac{v}{1000\;{\rm km\;s^{-1}}}\right)^{-1}
	 \;h_{50}^{-1/2},
\end{equation}
where (also hereafter) $H_0=50h_{50}$ km s$^{-1}$ Mpc$^{-1}$ is the Hubble
constant. The timescale of luminosity variability of a source due to
microlensing depends on the
mass of the microlens as $\sim(M/M_{\odot})^{1/2}$.
It turns out that the masses of the lensing objects can be determined by
monitoring the variations of the apparent luminosities of some sources
if the distances of lenses and sources as well as the transverse speed of the
lens $v$ are known. Note, however, that this procedure is actually restricted
by the sensitivities of the microlensing observations,
which  cannot include
the events with timescales lasting both shorter than the observing periods
of sampling and longer than the observing coverage.   The present observations
are then sensitive to those events with timescales
ranging from $\sim10$ minutes to $\sim10$ years.

The convincing evidence that a distant source is microlensed
has been found so far only in   QSO 2237+0305 ($z_s=1.695$)
associated with a foreground spiral galaxy ($z_d=0.0394$). This lens system
was discovered by Huchra et al. in 1985, in which the quasar appears to be
coincident with the nucleus of the galaxy. The subsequent observations
(De Robertis and Yee, 1988; Yee, 1988; Schneider et al., 1988) show that this
quasar actually consists of four components with the maximum separation of
$1.8^{\prime\prime}$
(see Figure 8), which is also referred to as the Einstein cross.
%@@@@@@@@@@@@@@@@@@@@@@@@@@@@@@@@@@@@@@@@@@@@@@@@@@@@@@@@@@@@@@@@@@@@@@
%                          FIGURE 8
\begin{figure}
 \vspace{1cm}
 \caption{Microlensing event QSO 2237+0305 (Figure courtesy of R. Stabell)}
\end{figure}
%@@@@@@@@@@@@@@@@@@@@@@@@@@@@@@@@@@@@@@@@@@@@@@@@@@@@@@@@@@@@@@@@@@@@@@
%@@@@@@@@@@@@@@@@@@@@@@@@@@@@@@@@@@@@@@@@@@@@@@@@@@@@@@@@@@@@@@@@@@@@@@
%                          FIGURE 9
\begin{figure}
 \vspace{1cm}
 \caption{Light curves of the four components of QSO 2237+0305. The data
	  are taken from Cumming and De Robertis (1995) and the error
	  associated with each datum point is typically 0.05 magnitude.}
\end{figure}
%@@@@@@@@@@@@@@@@@@@@@@@@@@@@@@@@@@@@@@@@@@@@@@@@@@@@@@@@@@@@@@@@@@@@@@
The photometric monitoring of QSO 2237+0305 has been made for several
years and the brightness variations of each component are shown in Figure 9.
Overall, the light curves of four components show no apparent correlations,
indicating that their variations must be due to
the microlensing magnification by the compact objects
and their relative motions in the spiral galaxy. Note that
the time delay between the components from the microlensing
is estimated to be about one day.
To be specific, a sharp variation of at least 0.2 magnitude in
the component A was detected within 26 days during 1988-1989
(Irwin et al., 1989; Corrigan et al., 1991), while the other components
didn't exhibit a similar feature. Another strong microlensing occurred
in component B in 1991: the brightness of B is magnified by a factor of
about 0.5 magnitude (Yee and De Robertis, 1992; Racine, 1992). As a
comparison, components C and D appear to be relatively stable.

%----------------------------------------------------------------------
\subsection{MACHO searches}

The inferred masses from the rotation curves of galaxies, including
our Galaxy, are an order of magnitude larger than their luminous masses,
implying that galaxies are embedded in  invisible massive  halos
(see Ashman 1992  for a recent review). The nature of the dark matter
in the halos of galaxies is still unknown today.
Basically, two kinds of candidates have been proposed: WIMPs
(weakly interacting massive particles such as axions and supersymmetric
neutralinos) and MACHOs (massive astrophysical compact halo objects
such as brown dwarfs, low mass stars and black holes).  WIMPs
are some kinds of unknown non-baryonic matter which may dominate the
Universe according to the standard inflation cosmological model and
the Big Bang primordial nucleosynthesis, especially
if the new measurement of the deuterium abundance in a high redshift
primordial hydrogen cloud (Songalia et al., 1994; Carswell et al., 1994)
is confirmed, whilst astronomers might favour MACHOs in the sense
that MACHOs may cause the observational effect -- microlensing.

If the relative velocity of the microlens $M$ (or the source)
transverse to the line of sight  is $v$, the
variability of the total magnification of a pointlike source follows
eq.(24). Figure 10 shows the light curve of a background source passing
%@@@@@@@@@@@@@@@@@@@@@@@@@@@@@@@@@@@@@@@@@@@@@@@@@@@@@@@@@@@@@@@@@@@@@@
%                          FIGURE 10
\begin{figure}
 \vspace{1cm}
 \caption{Light curve of a pointlike source passing behind a
          microlens (marked by ``$\bullet$") with a relative velocity of $v$.
	  The impact distance is $0.25a_E$ and $a_E$ is shown by
          the dotted lines.}
\end{figure}
%@@@@@@@@@@@@@@@@@@@@@@@@@@@@@@@@@@@@@@@@@@@@@@@@@@@@@@@@@@@@@@@@@@@@@@
near a microlens with impact parameter of $0.25a_E$. The main
characteristic features of the microlensing light curve are the
achromaticity and time-symmetry around the point of maximum magnification,
which can then be distinguished from the known variable star phenomena.

The probability that a source is gravitationally lensed is described by the
so-called {\it optical depth} ($\tau$).
The optical depth to microlensing without involving the
lenses at cosmological distance
is simply the number of microlenses or MACHOs inside the ``microlensing
tube" which has a cross-section of $\pi a_E^2$:
%45
\begin{equation}
\tau=\int_{0}^{D_s}\; \pi a_E^2\;n(D_d)\;dD_d,
\end{equation}
where $n(D_d)$ is the number density of the MACHOs at distance $D_d$.
Note that within the microlensing tube, $\mu>1.34$ or $|\Delta m|>0.32$.
Suppose that the Galactic halo is composed of MACHOs, their density profile
can be estimated from the rotational velocity $v_G$ which is about $220$ km/s
at the position of the Sun.
The original work by Paczy\'nski (1986) used a
SIS for the massive halo, and the subsequent work by Griest (1991) modified
SIS by introducing a core radius (ISC) which actually does not provide
any significant difference from SIS in the calculation of $\tau$. Employing
ISC for the MACHO distribution of the Galactic halo in eq.(45) yields
%46
\begin{equation}
\tau_G= \tau_0\int_{0}^{x_s}\frac{x(x_s-x)dx}
{x_s(1+x_c^2-2x\cos\alpha+x^2)},
\end{equation}
and
$$
\tau_0  =  \left(\frac{v_G}{c}\right)^2\frac{1}{1-\frac{r_c}{D_h}
\arctan\frac{D_h}{r_c}},
$$
where $r_c$ is the core radius of ISC,
$\alpha$ is the angle between the line of sight to the source
and the direction to the Galactic center, and $D_h$ is the
extent of the Galactic halo. All the distances are measured in unit of
$R_{GC}$, the distance to the Galactic center ($\approx8.5$ kpc),
so that $x=D_d/R_{GC}$, $x_s=D_s/R_{GC}$ and $x_c=r_c/D_{GC}$.

Now consider a star in our neighbour galaxy, the LMC, as the target.
The light rays from a star of LMC reach
the observer by passing through the LMC halo, the LMC disk and finally,
the Galactic halo. MACHOs of both the LMC halo/disk and the Galactic halo
are able to gravitationally magnify the apparent luminosity of the star
in the LMC. The contribution of the LMC halo to the microlensing optical
depth can be computed in a similar way to that of the Galactic halo
%47
\begin{equation}
\tau_{LMC,halo}  =  \tau_0\int_{x_h}^{x_s}\frac{x(x_s-x)dx}
{x_s(1+x_c^2-2x\cos\beta+x^2)},
\end{equation}
and
$$
\tau_0  =  \left(\frac{v_{LMC}}{c}\right)^2\frac{1}{1-\frac{r_c}{R_h}
\arctan\frac{R_h}{r_c}}.
$$
Here $r_c$ is the core radius of the LMC halo described by ISC,
$\beta$ is the angle between the line of sight to the star
and the center of the LMC, and $R_h$ is the
extent of the LMC halo. The distances in eq.(47) are measured in unit of
$D_{c}$, the distance to the LMC center ($\approx50.6$ kpc),
so that $x=D_d/D_c$, $x_s=D_s/D_c$ and $x_c=r_c/D_c$.
The LMC disk can be modeled by
an isothermal self-gravitating disk having density
(van der Kruit and Searle, 1981)
$n=n_0\exp(-R/h){\rm sech}^2(z/z_0)$, with the scale length $h$ in the radial
direction and the scale height $z_0$ in the $z$-direction.
The central number density $n_0$ relates with the maximum rotational
velocity ($v_m$) through (Freeman, 1970) $n_0=(1/4\pi MGhz_0)(v_m/0.62)^2$,
in which $M$ is the mass of the MACHO.
Furthermore, the inclination of the LMC disk can be simply taken to be
$0^o$, i.e., a face-on distribution of the disk matter. The optical
depth to microlensing by the disk is then
%48
\begin{equation}
\tau_{LMC,disk}  =  \tau_0\int_{0}^{x_s}\frac{x(x_s-x)}{x_s}
e^{-\frac{x\sin\beta}{(h/D_c)}}
{\rm sech}^2\left(\frac{1-x\cos\beta}{(z_0/D_c)}\right)dx,
\end{equation}
and
$$
\tau_0  =  \left(\frac{v_{m}}{0.62c}\right)^2
\left(\frac{D_c^2}{hz_0}\right).
$$

The contributions of the Galactic halo, the LMC halo and
the LMC disk to the microlensing optical depth for the stars of the LMC
are plotted in Figure 11.
%@@@@@@@@@@@@@@@@@@@@@@@@@@@@@@@@@@@@@@@@@@@@@@@@@@@@@@@@@@@@@@@@@@@@@@
%                          FIGURE 11
\begin{figure}
 \vspace{1cm}
 \caption{Optical depth to microlensing of the stars of the LMC by
          the Galactic halo, the LMC halo and the LMC disk.
          The extent of the LMC halo is taken to be $15^o$ (Schommer
          et al., 1992) and the LMC rotational velocity and $v_m$ are
          both taken to be 79 km/s.}
\end{figure}
%@@@@@@@@@@@@@@@@@@@@@@@@@@@@@@@@@@@@@@@@@@@@@@@@@@@@@@@@@@@@@@@@@@@@@@
The halo of our Galaxy gives rise to a nearly constant optical depth crossing
the LMC disk, $\tau_G=5\times10^{-7}$, while the LMC halo/disk provide
an optical depth depending sharply on the positions of the stars, which
arises from the fact that only the foreground MACHOs of the LMC are able
to act as lenses for the stars of the LMC itself.   For the stars near
the LMC center ($\beta\leq0^o.5$), $\tau_{\rm LMC, halo}\geq2\times 10^{-7}$
and $\tau_{\rm LMC,disk}\geq3\times 10^{-8}$.  Therefore, if the halo of
our Galaxy and of the LMC are composed of MACHOs, several million
stars should be monitored for the discovery of the microlensing events
in the LMC.

Following the proposal of Paczy\'nski (1986), the EROS (Exp\'erience
de Recherche d'Objets Sombres) and the MACHO collaboration commenced
in 1990 their searches for microlensing events of the LMC by monitoring
the brightness of a few million stars in the LMC, and the OGLE (Optical
Gravitational lensing Experiment) began in 1992 to conduct a similar search
in the direction of the Galactic bulge. Three groups announced their
discoveries at almost the same time in the autumn of 1993:
Three events were detected in the LMC (Alcock et al. 1993; Aubourg et al. 1993)
and six were seen in the Galactic bulge (Udalski et al., 1993).
During the writing process of this article, the total ``local" microlensing
events have grown  to $\sim70$. These include more than 10 events found by
the OGLE collaboration (Udalski et al., 1994a; 1994b; 1995) and
more than 40 events by the MACHO collaboration
(Bennett et al., 1994; Alcock et al., 1995a) in the Galactic bulge, and
3 events by the MACHO collaboration (Alcock et al., 1995b)
and 2 events by the EROS team in the LMC [Note that the EROS $n^o$2
candidate may be an eclipsing binary system rather than a microlensing
event (Ansari et al., 1995)]. Figure 12
shows the light curves of, and the lensing model fits to the MACHO
microlensing events and Table 2 summarizes the properties of
the 5 microlensing candidates of the LMC, which can be  regarded
as the representatives of all the reported microlensing candidates.
%@@@@@@@@@@@@@@@@@@@@@@@@@@@@@@@@@@@@@@@@@@@@@@@@@@@@@@@@@@@@@@@@@@@@@@
%                          FIGURE 12
\begin{figure}
 \vspace{1cm}
 \caption{The observed light curves of the three MACHO microlensing
	   candidates and the best-fit theoretical lensing model.
	   [From Alcock et al. (1995b)]}
\end{figure}
%@@@@@@@@@@@@@@@@@@@@@@@@@@@@@@@@@@@@@@@@@@@@@@@@@@@@@@@@@@@@@@@@@@@@@@
%********************************* TABLE 2 ***************************
\begin{small}
   \begin{table}
      \caption{Properties of the microlensing candidates
               of the LMC}
         \label{ }
      \[
\begin{array}{|c|c|c|c|c|c|}
\hline
{\rm candidate} & {\rm RA(2000)} & {\rm Dec(2000)} &{\rm media\; magnitude} (V)
&
 \mu_{max} & T\; ({\rm days})\\
\hline
{\rm MACHO}\; n^o1 & 05\,15\,44.5 & -68\,48\,00& 19.6 & 7.20 & 34.8\\
{\rm MACHO}\; n^o2 & 05\,22\,57.0 & -70\,33\,14& 20.7 & 1.99 & 19.8\\
{\rm MACHO}\; n^o3 & 05\,29\,37.4 & -70\,06\,01& 19.4 & 1.52 & 28.2\\
{\rm EROS}\; n^o1  & 05\,26\,36 & -70\,57\,37& 19.0 & 2.5 & 27\\
{\rm EROS}\; n^o2  & 05\,06\,06 & -65\,58\,34& 19.3 & 3.0 & 30\\
\hline
\end{array}
   \]
   \end{table}
\end{small}
%********************************* TABLE 2 ***************************

If these events are indeed generated
by the microlensing of the MACHOs along the light of sight rather than a
new kind of variable stars, one can estimate the mass of the MACHOs
using the event duration  $T$ (i.e., the Einstein ring crossing time)
and the maximum magnification ${\mu}_{max}$.
At the point where the background star enters into the microlensing tube
(denoted by subscript ``$_{min}$"), eq.(24) reads
%49
\begin{equation}
{\mu}_{min}=\frac{u_{min}^2+2}{u_{min}\sqrt{u_{min}^2+4}}=1.34,
\end{equation}
in which  $u_{min}=1$, while at the maximum magnification
(denoted by subscript ``$_T$")
%50
\begin{equation}
{\mu}_T=\frac{u_T^2+2}{u_T\sqrt{u_T^2+4}}.
\end{equation}
Alternatively, $u_T$ is related with $u_{min}$ through geometrical relation
%51
\begin{equation}
u_{min}^2-u_T^2=\left(\frac{vT}{2a_E}\right)^2,
\end{equation}
where $v$ is the relative velocity of the star or the microlens.
These three equations give rise to the mass of microlens
%52
\begin{equation}
M=\frac{c^2}{32G}\;\frac{D_s}{D_dD_{ds}}\;
\frac{v^2T^2}{\frac{{\mu}_{min}}{\sqrt{{\mu}_{min}^2-1}}-
              \frac{{\mu}_T}{\sqrt{{\mu}_T^2-1}}}.
\end{equation}
If the MACHOs of Galactic halo are responsible for the observed events,
we can take $D_d=10$ kpc and $v=220$ km/s for a numerical estimate.
Utilizing the observed microlensing event duration of typically $T\sim30$ days
and the maximum magnification of a few leads to  $M\sim0.1M_{\odot}$,
i.e., sub-solar objects in the Galactic halo are likely to be the
deflectors for the microlensing events of the LMC.

However, one cannot conclude from the presently detected  microlensing
events in the LMC that the halos of the galaxies are dominated
by $\sim0.1$ solar mass objects. In fact, the positions $D_d$ and
relative velocities $v$ of traverse motion of the lensing objects are
two unknown factors in the determination of masses of  MACHOs.
Unless a statistical sample of microlensing events in the LMC is
completed, it is in principle impossible to draw a decisive conclusion
about the masses of the MACHOs in the Galactic halo.
Gould (1994) argued that the observed optical depth toward the
LMC center in the MACHO collaboration is only $7-9\times10^{-8}$,
much less than that expected from the Galactic halo made of MACHOs.
Using the fact that 3 events were detected among 9.5 million monitored
stars in LMC for 1.1 years, the MACHO Collaboration
(Alcock et al., 1995a,c)
has recently reached a similar microlensing optical depth of
$8.8^{+7}_{-5}\times10^{-8}$, nearly an order of magnitude lower
than the expected optical depth of $\sim 5\times 10^{-7}$.
So, the halo of our Galaxy may have not been detected at all. If so,
the microlensing events seen by EROS and MACHO collaboration
may have arisen from the stars of the LMC disk (Wu, 1994b; Sahu, 1994)
and  self-lensing by a  stellar disk remains to be an interesting
model for further investigation (Gould, 1995).\\

%----------------------------------------------------------------------
\subsection{Simulations and observations}

Cosmological compact objects either bounded in galaxies or distributed
randomly in the Universe are capable of magnifying temporarily the
background sources like quasars, AGNs, etc., resulting in  variations
of their apparent luminosities.  Besides the significant difference of
timescales between  the cosmological microlenses and the
local ones like those in the Galactic halo and in the LMC [see eqs.(43)
and (44)] which we have discussed in the above subsection,
the optical depth to microlensing arising from the lenses at
cosmological distance
may be a few orders of magnitude larger than the local optical depth,
depending on the content of compact objects of the Universe.
For example, the optical depth to microlensing for a distant source at
redshift $z_s=3\sim4$ can be of order of unity if the Universe is composed
of compact objects (see Figure 17).  So, one now needs to deal with the
problem that a background source is simultaneously microlensed by
$n$--pointlike masses.

Basically, for an ensemble of compact objects as microlenses which are
often assumed to be on a single lens plane, the total magnification of
a pointlike source at the position $\mbox{\boldmath $\beta$}$ is the sum of
the magnification $\mu_i$ of each micro image at position
$\mbox{\boldmath $\theta_i$}$ on the lens plane
%53
\begin{equation}
\mu(\mbox{\boldmath $\beta$})=\displaystyle\sum_i\mu_i=
\displaystyle\sum_i\left| {\rm det} \frac{\partial\mbox{\boldmath $\beta$}}
{\partial \mbox{\boldmath $\theta_i$}}  \right|^{-1}.
\end{equation}
However, this straightforward method cannot be efficiently employed for
a computation of  the total magnification. In practice, one is unable
to find analytically all the micro images when the number of lenses is very
large, especially for  extended sources.
%@@@@@@@@@@@@@@@@@@@@@@@@@@@@@@@@@@@@@@@@@@@@@@@@@@@@@@@@@@@@@@@@@@@@@@
%                          FIGURE 13
\begin{figure}
 \vspace{1cm}
 \caption{($a$)Magnification patterns on the source plane by an ensemble of $N$
          (left) and $N-1$ (right) star-like microlenses  and
	  ($b$)four horizontal (left) and four vertical (right) light curves
	  for a source crossing the tracks marked with think black lines in
	  ($a$). The thick  and thin lines correspond to ($a$)-left and
	  ($b$)-right, respectively. (Figure courtesy of J. Wambsganss)}
\end{figure}
%@@@@@@@@@@@@@@@@@@@@@@@@@@@@@@@@@@@@@@@@@@@@@@@@@@@@@@@@@@@@@@@@@@@@@@
Many numerical techniques
have been  developed to deal with the problem of large
number of microlenses including   Monte-Carlo simulation (Young, 1981),
the ray-shooting method (Kayser, Refsdal and Stabell, 1986;
Schneider and Weiss, 1987),
the Fourier method (Katz, Balbus and Paczy\'nski, 1986),
the Markoff method (Deguchi and Watson, 1988),
and the parametric representation of caustics (Witt, 1990).
In particular, the inverse ray-shooting method has been widely used
in recent years in  microlensing simulations: Light rays are traced
backwards from the observer to the source plane, on which the magnification
pattern is represented by the intersection of the rays. So, the number density
of rays is proportional to the magnification.  A typical magnification
pattern produced by an ensemble of $N$ star-like microlenses is shown in
Figure 13. If a background source  traverses the magnification regions
due to either the motion of the source itself or the
velocity dispersion of the stars associated with the lens galaxies,
the apparent luminosity of the source would vary as a function of time.
To most observers, this kind of variability is something like a ``noise".
However, it should be noticed that the microlensing-induced variations
could be very dramatic sometimes, which may explain the
unusual features associated with some special objects.

It was noticed that the violently variable objects 0846+51W1 (quasar),
AO 0235+164 (BL Lac object) and PKS 0537-441 (blazar)  might be
the results of microlensing (Nottale, 1986; Stickel, Fried and K\"uhr,
1988a,b; 1989). It was even speculated that the variability of apparent
magnitude of quasars  are
partially due to microlensing rather than their intrinsic physical processes
(Peacock, 1986; Kayser, Refsdal and Stabell, 1986; Schneider and Weiss, 1987).
These arguments have
recently  been strengthened by Hawkins (1993), based on the analysis of
a complete sample of $\sim300$ quasars selected from their variability
over 17 years (Hawkins and V\'eron, 1993; hereafter HV).
Some typical light curves
from their sample are plotted in Figure 14 for two high redshift
($z_s=2$)  and two low redshift ($z_s=0.2$) quasars, respectively.
To investigate  whether these variabilities are intrinsic to quasars or
%@@@@@@@@@@@@@@@@@@@@@@@@@@@@@@@@@@@@@@@@@@@@@@@@@@@@@@@@@@@@@@@@@@@@@@
%                          FIGURE 14
\begin{figure}
 \vspace{1cm}
 \caption{The typical light curves of four variability-selected
	  quasars by HV. The left panels are high redshift quasars at
          $z_s\approx2$ and the right panels are low redshift ones
          at $z_s\approx0.2$.}
\end{figure}
%@@@@@@@@@@@@@@@@@@@@@@@@@@@@@@@@@@@@@@@@@@@@@@@@@@@@@@@@@@@@@@@@@@@@@@
due to gravitational lensing by  compact objects along the
lines of sight,  an analysis of the time-varying autocorrelation function
was made. It turns out that the timescale of quasar luminosity variations
decreases  with increasing redshift. This is inconsistent with the theory
that the expansion of the Universe should cause observed timescales to increase
linearly with $(1+z_s)$ due to time dilation. Therefore,
Hawkins concluded that
the quasar variabilities cannot be intrinsic to quasars themselves, and
gravitational lensing is the most possible cause, indicative of
the existence of a large number of compact objects up to $10\%$ of the
critical mass density of the Universe.
Nevertheless, this claim should be taken very cautiously in the sense
that the observed feature of  quasar variability increasing with their
redshift can also be interpreted as the result of cosmic evolution or
various observational limitations, e.g. the finite duration of the monitoring
campaign, the finite photometric sensitivity (Alexander, 1995) and
the observing wavelenght dependence arising from the accretion disk model
of quasar (Baganoff and Malkan, 1995).
Another interesting issue is the number deficit in HV,
as compared with the optically selected quasars (Boyle, Shanks and
Peterson, 1988; hereafter BSP) (see Figure 15). It remains worth
investigating whether the quasar number discrepancy in HV is related
to microlensing or to the observational methods.
%@@@@@@@@@@@@@@@@@@@@@@@@@@@@@@@@@@@@@@@@@@@@@@@@@@@@@@@@@@@@@@@@@@@@@@
%                          FIGURE 15
\begin{figure}
 \vspace{1cm}
 \caption{The number-magnitude relations for the optically selected
          quasars (open circles) (BSP) and the
          variability-selected ones (open triangles) (HV)}
\end{figure}
%@@@@@@@@@@@@@@@@@@@@@@@@@@@@@@@@@@@@@@@@@@@@@@@@@@@@@@@@@@@@@@@@@@@@@@
Finally, the most important aspect of studying cosmological microlensing is to
set constraints on the fraction of compact objects ($\Omega_c$) in the matter
density of the Universe by analyzing  complete samples of
variability-selected sources,
which is quite similar to the purpose of the ongoing MACHO
experiments in our Galaxy, although one cannot separate the
microlensing-induced variability from the variability intrinsic to sources.
For instances, using the HV sample, Schneider
(1993) obtained an upper limits of $\Omega_c<0.1$ for  compact objects
with masses ranging from $10^{-3}M_{\odot}$ to $3\times10^{-2}M_{\odot}$.
Dalcanton et al. (1994) have recently found $\Omega_c<0.1$ in the mass range
$0.01M_{\odot}$ -- $20M_{\odot}$, $\Omega_c<0.2$ for $0.001M_{\odot}$ --
$60M_{\odot}$ and $\Omega_c<1$ for $0.001M_{\odot}$ -- $300M_{\odot}$ by
comparing the distributions of the AGN and quasar equivalent widths of
emission lines at low and high redshifts.  It is expected that observations
of cosmological microlensing can set more stringent limits on $\Omega_c$ in
the next few years. \\

%----------------------------------------------------------------------
\subsection{Inhomogeneous Universe}

Astrophysical observations indicate that the Universe tends to be
locally inhomogeneous on scales less than $\sim100$ Mpc.
Since the early 1960's there have been many studies about the influence of
matter inhomogeneities on the propagation of light rays from  distant
sources, especially on the magnitude-redshift ($m\sim z$) relation
(Zel'dovich, 1964; Bertotti, 1966; Gunn, 1967; Kantowski, 1969;
Dyer and Roeder, 1972; 1973; Canizares, 1982; Nottale, 1982a,b; 1983;
Vietri and Ostriker, 1983;  Schneider and Weiss, 1988a,b; Isaacson
and Canizares; 1989; Wu, 1990b; 1992b; Kantowski, Vaughan and
Branch, 1995).  Many authors addressed the
question if the classical $m\sim z$ relation (Mattig, 1958)
should be modified, i.e., if the $m\sim z$ relation in a locally
inhomogeneous Universe differs from that in a standard Friedmann-Lemaitre
Universe.

In an inhomogeneous Universe the propagation of a bundle of
light rays is controlled by two different effects: A beam of light
traveling outside the mass clump would diverge faster because the matter
density in such a region is lower than the mean matter density of the
Friedmann-Lemaitre Universe, whereas a beam of light passing near the
clump would be sheared by the gravity of the clump, leading to
the convergence of light rays. Flux conservation requires that the
divergence due to the absence of matter inside the beam  be
balanced on average
by the convergence due to the gravitational effect of the
clump, so that the luminosity distances in both the inhomogeneous
Universe and the Friedmann-Lemaitre Universe remain statistically equal
if the size of the inhomogeneities is sufficiently small (typically, the size
of galaxy) (Weinberg, 1976).

Suppose that the Universe is uniformly filled by both the intergalactic
medium of density of $\tilde{\alpha}\rho$ and  matter clumps of density of
$(1-\tilde{\alpha})\rho$ so that the mean mass density is the same as
that in the Friedmann-Lemaitre Universe. $\tilde{\alpha}$ denotes the
fraction of the total mass density that is intergalactic.
$\tilde{\alpha}=1$ corresponds to a completely homogeneous Universe, i.e.,
the Friedmann-Lemaitre model, and $\tilde{\alpha}=0$ describes a completely
inhomogeneous Universe in which all the matter is concentrated into clumps.
For a beam of light rays with vertex at the observer ($z=0$)
propagating far away from any matter clumps, the
shearing effect of the beam can be neglected and the propagation of light is
determined by the optical scalar equation (Sachs, 1961)
%54
\begin{equation}
\begin{array}{l}
x(x-1)\frac{d^2\tilde{D}}{dx^2}+\left(\frac{7}{2}x-3\right)
\frac{d\tilde{D}}{dx}+\frac{3}{2}\tilde{\alpha} \tilde{D}=0;\\
x=\frac{\Omega}{\Omega-1}(1+z)
\end{array}
\end{equation}
in which we use  the angular diameter distance $\tilde{D}$ (in units of
$c/H_0$) as the variable (Dyer and Roeder, 1973),
%55
\begin{equation}
\Omega\equiv \frac{\rho}{\rho_c}
\end{equation}
is the cosmological mass density parameter and
%56
\begin{equation}
{\rho_c}\equiv \frac{3H_0^2}{8\pi G}
\end{equation}
is the critical mass density of the Universe.  The initial conditions
of eq.(54) can be conveniently chosen to be the values of angular diameter
distance and the local expansion of the Universe at the observer:
%57
\begin{equation}
\begin{array}{l}
\tilde{D}|_{z=0}=0;\\
\frac{d\tilde{D}}{dz}|_{z=0}=1.
\end{array}
\end{equation}
Under these initial conditions the solution to eq.(54)
with $\tilde{\alpha}=1$ gives the well-known angular diameter
distance in the Friedmann-Lemaitre Universe
%58
\begin{equation}
\tilde{D}=\frac{2[z\Omega+(\Omega-2)(-1+\sqrt{1+\Omega z})]}{\Omega^2(1+z)^2}.
\end{equation}
In the case of $\tilde{\alpha}=0$ eq.(54) reduces to (Dyer and Roeder, 1972)
%59
\begin{equation}
\tilde{D}=\int_0^{z}\frac{dz}{(1+z)^3\sqrt{1+\Omega z}}.
\end{equation}
In particular, the solution to eq.(54) for a flat Universe of $\Omega=1$ is
%60
\begin{equation}
\tilde{D}=\frac{2}{\tilde{\beta}}(1+z)^{(\tilde{\beta}-5)/4}
\left[1-(1+z)^{-\tilde{\beta}/2}\right]
\end{equation}
in which $\tilde{\beta}=\sqrt{25-24\tilde{\alpha}}$. Seitz and Schneider (1994)
have recently shown that a general solution to eq.(54)
can be found for any values of $\Omega$
and $\tilde{\alpha}$ by transforming
eq.(54) into the Legendre differential equation.

Luminosity distances in the Friedmann-Lemaitre Universe and in the
inhomogeneous Universe can be denoted by  $D_{L0}$ and $D_L$,  respectively,
and relate with angular diameter distances by multiplying a factor
of $(1+z)^2$. Their difference reflects the divergence of the light
propagation in the Universe, whilst
the convergence can be described by gravitational lensing effect
when a beam of light rays passes near the clumps. Hence, the parameter
%61
\begin{equation}
\Delta m=5\log \frac{D_L}{D_{L0}}-2.5\log {\mu}
\end{equation}
indicates a deviation of the actually observed apparent magnitude of a source
from its theoretically expected value in a completely homogeneous Universe.
Note that $\Delta m$ may have relatively large variations,
i.e., a distant source may dramatically change
its apparent luminosity. This arises because
the magnification ${\mu}$ can vary in principle from unity to infinity,
depending on the distance of the line of sight to the background source from
the mass clumps.  However, energy conservation requires
that luminosity distances in a homogeneous Universe and a clumpy Universe
should be statistically equal, which reads (Ehlers and Schneider, 1986)
%62
\begin{equation}
\langle {\mu} \rangle=\left(\frac{D_L}{D_{L0}}\right)^2.
\end{equation}
Thus, the mean magnification correction is
%63
\begin{equation}
\langle \Delta m\rangle =2.5[\log \langle {\mu}\rangle -
\langle \log {\mu}\rangle].
\end{equation}

The probability that a source at $z_s$ is gravitationally magnified by
a factor of ${\mu}$ due to  pointlike lenses within
($z_d$, $z_d+dz_d$) is
%64
\begin{equation}
dp=n_d\;(\pi\theta^2D_d^2)\;(dr_{prop}/dz_d)dz_d
\end{equation}
where $n_d$ is the number density of the lenses, $dr_{prop}$, the
differential proper distance around the lens at $z_d$,
and $\pi\theta^2D_d^2$ is the lensing
cross-section which is given by the lensing equation eq.(24):
%65
\begin{equation}
\pi\theta^2=2\pi \theta_E^2\;\left(\frac{{\mu}}{\sqrt{{\mu}^2-1}}-1\right).
\end{equation}
Assuming a uniform distribution of pointlike lenses in the Universe,
i.e. $n_d=(1+z_d)^3n_{d0}$,
and defining the mass density parameter of the lenses as
%66
\begin{equation}
(1-\tilde{\alpha})\Omega\equiv
\frac{n_{d0}M}{\rho_c}=\frac{8\pi G M n_{d0}}{3H_0^2},
\end{equation}
we have the total probability for a source at $z_s$ to be magnified by
a factor of $\mu$ due to  foreground compact objects
%67
\begin{equation}
P_1({\mu})=3\Omega(1-\tilde{\alpha})\frac{H_0}{c}
\left(\frac{{\mu}}{\sqrt{{\mu}^2-1}}-1\right)\;
\int_0^{z_s}\frac{D_dD_{ds}}{D_s}\frac{1+z_d}{\sqrt{1+\Omega z_d}}dz_d.
\end{equation}
Fortunately, this expression can be separated into two parts
%68
\begin{equation}
P_1({\mu})=f_1({\mu})\;\tau
\end{equation}
where
%69
\begin{equation}
f_1({\mu})=2\left(\frac{{\mu}}{\sqrt{{\mu}^2-1}}-1\right);\\
\end{equation}
%70
\begin{equation}
\tau=\frac{3}{2}\Omega(1-\tilde{\alpha})\frac{H_0}{c}
\int_0^{z_s}\frac{D_dD_{ds}}{D_s}\frac{1+z_d}{\sqrt{1+\Omega z_d}}dz_d.
\end{equation}
$\tau$ is the optical depth to gravitational lensing in terms of the
definition of eq.(45), which describes the total number of lenses enclosed
within the Einstein ring along the line of sight to the source.
Figure 16 shows the ``maximum" optical depth contributed by
 pointlike lenses that compose all the matter of the Universe,
i.e., $\tilde{\alpha}=0$.
%@@@@@@@@@@@@@@@@@@@@@@@@@@@@@@@@@@@@@@@@@@@@@@@@@@@@@@@@@@@@@@@@@@@@@@
%                          FIGURE 16
\begin{figure}
 \vspace{1cm}
 \caption{The optical depth to gravitational lensing by the pointlike
          lenses in a completely inhomogeneous Universe
          ($\tilde{\alpha}=0$ and $\Omega=1$).}
\end{figure}
%@@@@@@@@@@@@@@@@@@@@@@@@@@@@@@@@@@@@@@@@@@@@@@@@@@@@@@@@@@@@@@@@@@@@@@
Note that eq.(70) is valid for computation of the optical depth to
gravitational lensing by various compact objects with different masses.
Meanwhile, $\tau$ is  an indicator of the significance of multiple lenses.
In the case of $\tilde{\alpha}=0$ and $\Omega_0=1$
(Figure 16), a source with $z_s>3$ may be lensed by more than one lensing
object ($\tau\sim1$).  Therefore, multiple lenses need to be taken into
account for the high-redshift sources if a relatively large amount of
matter of the Universe is concentrated into compact objects.

Multiple lenses fall into two classes: geometrically-thin multiple lenses
located on a single lens plane, as was discussed in the above
subsection, and spatially discrete multiple lenses which are distributed on
different lens planes along the line of sight.  For the multiple lens
plane deflections, the gravitational
lensing equation can be formally written as
(Blandford and Narayan, 1986; Schneider, Ehlers and Falco, 1992)
%71,72
\begin{eqnarray}
\mbox{\boldmath $\eta$}=\frac{D_s}{D_1}\mbox{\boldmath $\xi$}_1
-\displaystyle\sum_{i=1}^N D_{is} \mbox{\boldmath $\alpha$}_i
(\mbox{\boldmath $\xi$}_i),\\
\mbox{\boldmath $\xi$}_j=\frac{D_j}{D_i}\mbox{\boldmath $\xi$}_1
-\displaystyle\sum_{i=1}^{j-1} D_{ij} \mbox{\boldmath $\alpha$}_i
(\mbox{\boldmath $\xi$}_i),
\end{eqnarray}
where $\mbox{\boldmath $\eta$}$ is the position of the source  in source
plane,  $\mbox{\boldmath $\alpha$}_i(\mbox{\boldmath $\xi$}_i)$
is the deflection angle of the light ray with
the impact distance  $\mbox{\boldmath $\xi$}_i$ by
the deflectors in the $i$-th lens plane, $D_{ij}$ denotes the angular
diameter distance from the $i$-th lens plane to the $j$-th lens plane and
$D_i$, from the observer to the $i$-th lens plane (Note that $j=s$ refers
to the source).  The total magnification of the source luminosity is
finally given by the inverse of the determinant of the magnification matrix
which relates to the Jacobian matrices  of the mapping from the $(i-1)$-th
lens plane to the $i$-th lens plane ($1<i<N$). Both  numerical techniques
(e.g. Schneider and Weiss, 1988a,b) and analytical methods
(Seitz and Schneider, 1992, 1994) have been employed in the determination of
the magnification factor as well as the magnification probability
of a distant source by the deflectors in multiple lens planes, although
the procedures turn to be relatively complex.

In the practical computation of magnification probability by multiple lenses,
some approximations are often employed in order to avoid the above complexity.
The fact that the probability
$P_1({\mu})$ by a single lens factorizes into two independent parts in eq.(68)
leads to the speculation that the probability
$P_n({\mu})$ for a distant source
by $n$ lenses be written as a product (Wu, 1990a)
%73
\begin{equation}
P_n({\mu})=f_n({\mu})\;g_n(\tau),
\end{equation}
i.e., the ${\mu}$ variable separates from the $\tau$ variable.
$g_n(\tau)$ is a function that describes the probability of a background
source being lensed by  $n$ foreground objects.
Obviously, $g_n(\tau)$ follows a Poisson distribution
%74
\begin{equation}
g_n(\tau)=\frac{\tau^n}{n!}\;e^{-\tau}.
\end{equation}
Therefore, the magnification distribution function can be obtained by
summing up $P_n({\mu})$ over all $n$
%75
\begin{equation}
P({\mu})=\displaystyle \sum_{n=0}^{\infty}\;
\frac{\tau^n e^{-\tau}}{n!}\;f_n({\mu}).
\end{equation}
Unfortunately, no exact expressions have been established for
$P_{\geq2}({\mu})$.
An oversimple assumption is that the total magnification ${\mu}$ is the
product of the individual magnification ${\mu}_i$ so that $P({\mu})$
is the convolution of $P_i({\mu})$ of each lens (Canizares, 1982;
Vietri and Ostriker, 1983; Peacock, 1986; Isaacson and Canizares, 1989;
Schneider, 1993; Pei, 1993). Nevertheless, a justification for
this assumption is rather hard, and it seems that
this multiplication method may be questionable (Wu, 1990a).

Finally, in a single lensing approximation the mean correction of the apparent
magnitude [eq.(63)] to the Mattig's relation is (Wu, 1990b,1992b)
%76
\begin{equation}
\langle \Delta m\rangle =1.16\tau+O(\tau^2)\approx 0.58(1-\tilde{\alpha})
(\Omega/2)z_s^2+\cdot\cdot\cdot,
\end{equation}
in comparison with the correction found by Kantowski (1969) and
Dyer and Roeder (1974) based on the Swiss-cheese model for
the inhomogeneities of the Universe:
%77
\begin{equation}
\langle \Delta m\rangle\approx 1.086(1-\tilde{\alpha})
(\Omega/2)z_s^2+\cdot\cdot\cdot.
\end{equation}
These formulae can be used to statistically estimate the modification to the
classical magnitude-redshift relation due to the matter inhomogeneities,
 and the quantity depends sharply on the matter content of
compact objects $(1-\tilde{\alpha})\Omega$  in the Universe.\\

\bigskip
%@@@@@@@@@@@@@@@@@@@@@@@@@@@@@@@@@@@@@@@@@@@@@@@@@@@@@@@@@@@@@@@@@@@@@@@
\bigskip
\section{GALAXIES AND MULTIPLE IMAGES}

%-------------------------------------------------------------------
\subsection{Multiply-imaged quasars}

The most significant feature of gravitational lensing is the multiple imaging
of the lensed source. The first gravitational lens system was discovered
in 1979 (Walsh, Carswell and Weymann, 1979) during the identification of
the optical counterpart of a radio source. This famous system 0957+561
consists of two quasar images
(A and B) at the same redshift of $z_s=1.41$ and with
separation of $6''.1$. A galaxy at $z_d=0.36$ was soon detected at the
position near the image B (Adams and Boroson, 1979; Young et al., 1980;
Stockton, 1980), which is believed to be the main deflector with
mass of $\sim10^{12}M_{\odot}$ (see Figure 17).  VLBI  observation of
0957+561A,B (e.g. Gorenstein et al., 1988; Garret et al., 1994) strongly
%@@@@@@@@@@@@@@@@@@@@@@@@@@@@@@@@@@@@@@@@@@@@@@@@@@@@@@@@@@@@@@@@@@@@@@
%                          FIGURE 17
\begin{figure}
 \vspace{1cm}
 \caption{Gravitational lens system QSO 0957+561A,B. [from E. Schild
          (1991)]}
\end{figure}
%@@@@@@@@@@@@@@@@@@@@@@@@@@@@@@@@@@@@@@@@@@@@@@@@@@@@@@@@@@@@@@@@@@@@@@
%@@@@@@@@@@@@@@@@@@@@@@@@@@@@@@@@@@@@@@@@@@@@@@@@@@@@@@@@@@@@@@@@@@@@@@
%                          FIGURE 18
\begin{figure}
 \vspace{1cm}
 \caption{VLBI observation of the gravitational lens system QSO 0957+561A,B.
         [from Garrett. et al. (1994)]}
\end{figure}
%@@@@@@@@@@@@@@@@@@@@@@@@@@@@@@@@@@@@@@@@@@@@@@@@@@@@@@@@@@@@@@@@@@@@@@
confirmed the lensing origin of QSO 0957+561A,B  by revealing the same
radio morphology (one compact core and three jets) of the two images
(Figure 18).

The criteria for the determination of a gravitationally lensed quasar
system are
(1)multiple images, (2)similar spectra and the same redshifts of the
images and (3)the detection of intervening galaxies as lenses. A
list of the accepted and the proposed lensed quasar systems compiled by
Surdej and Soucail in 1993 is updated in Table 3.
The proposed rather than confirmed cases arise mainly from the fact
that no corresponding deflectors have been found to be
responsible for the multiple images.  The absence of the lensing galaxies
in the direct imaging centered on some of the multiple quasars even with HST
is a well-known puzzle in gravitational lensing. It remains unclear today if
one should really reject these lensing candidates  in which the ``dark" lensing
galaxies are apparently missing. The recent detections of a very faint galaxy
($B=25.0$) and the faint cluster of galaxies at $z>1$ (Mellier et al., 1994)
associated with the double quasar 2345+007 (Fischer et al., 1994)
may be promising for the future searches of the deflectors in other
lensing candidate systems.

%********************** TABLE 3 **********************************
   \begin{table}
      \caption{Gravitationally-Lensed Multiple Quasars}
         \label{ }
      \vspace{1cm}
   \end{table}

%-------------------------------------------------------------------

\subsection{Multiply-imaged radio sources and radio rings}

Radio observation turns out to be an efficient way of finding multiple
images  due to its consistently high
dynamical range and resolution of the maps. Indeed,
the VLBI observation of double quasar 0957+561A,B (Figure 18) played an
important role in the confirmation of their lensing origin
(Gorenstein et al., 1988).  Today, with the completeness of a large
lens survey at the Very Large Array, the Cosmic Lens All-Sky Survey would
provide a large number of lensing candidates for both modeling of the
lensing systems and statistical study. The recent discoveries of a double lens
images 1600+434 (Jackson et al., 1995) and a quadruple lens system 1608+656
(Myers et al., 1995; Snellen et al., 1995)
have marked the success of this survey.

Another success of radio observation in gravitational lensing is the
detection of the Einstein ring.
As was illustrated in Figure 2, when the background source lies in  a
position behind the foreground lensing object, the multiple images
merge into a ring-like image, i.e. the Einstein ring.
The first radio ring (Figure 19) was discovered by
Hewitt et al.  in 1988, which is the image of a radio lobe at
$z_s=1.13$ lying perfectly behind a foreground galaxy at $z_d=0.85$.
%@@@@@@@@@@@@@@@@@@@@@@@@@@@@@@@@@@@@@@@@@@@@@@@@@@@@@@@@@@@@@@@@@@@@@@
%                          FIGURE 19
\begin{figure}
 \vspace{1cm}
 \caption{15 GHz  image of the radio Einstein ring MG1131+0456
          [from J. Hewitt et al. (1988)]}
\end{figure}
%@@@@@@@@@@@@@@@@@@@@@@@@@@@@@@@@@@@@@@@@@@@@@@@@@@@@@@@@@@@@@@@@@@@@@@
Six more radio rings have been so far observed:
MG $1654+1346$ (Langston et al., 1989), PKS $1830-211$ (Rao and Subrahmanyan,
1988; Jauncey et al., 1991), MG $1549+3047$,
MG $0751+2716$, B $1938+666$ (Leh\'ar et al., 1993),
and B $0218+357$ A$-$B (Patnaik et al., 1993).
Besides constraining the mass
profile in the lensing galaxies from modeling of the radio rings, the
measurement of time delay in the radio ring would be of great interest
for the determination of the Hubble constant (see section 3.4).

%-------------------------------------------------------------------
\subsection{Galaxies as Lenses}

Galaxies are found to be the main deflectors for the $\sim10$ confirmed
gravitationally-lensed multiple quasars in Table 3. Modeling each lens
system based on the observational data has reproduced
quite well the observed multiple images of quasars.
Actually, this procedure is no more than  to solve the lensing
equation for different gravitational potentials, which has been extensively
discussed in section 1.2.
The most important issue in the study of multiply-imaged quasars,
however, is the statistical properties, which raises the question if there
are enough massive galaxies in the Universe to be responsible for the observed
events of gravitationally-lensed quasars.  Recall that the lensing galaxies
have not been detected today in some of the proposed lens candidates.
Moreover, statistical lensing of galaxies provides the information on how
many multiply-imaged quasars would be expected to observe over the sky. By
comparing the theoretical statistical predictions with the observations of
multiple quasars, one can also set useful constraints on the mass density
of the lensing objects in the Universe (e.g.,  Hewitt et al., 1986;
Claeskens et al., 1993; Surdej et al., 1995)
and  the cosmological constant $\Lambda$ (e.g.,
Turner, 1990; Fukugita et al., 1992; Sasaki and Takahara, 1993; Rix et al.,
1994; Kochanek, 1992;1993a,c;1995).

Adopting the simplest matter distribution, SIS, for the lensing galaxy,
one can write the lensing cross-section from eq.(27) to be
%78
\begin{equation}
\pi\theta^2=\frac{\pi \theta_E^2}{({\mu}-1)^2}.
\end{equation}
Hence, the probability that a source at $z_s$ is magnified by a factor of
greater than ${\mu}$ due to an ensemble of galaxies within redshift $dz_d$
of $z_d$ is
%79
\begin{equation}
dp=F \;\left(\frac{\tilde{D}_d \tilde{D}_{ds}}{\tilde{D}_s}\right)^2
    \frac{(1+z_d)}{\sqrt{1+\Omega z_d}}
    dz_d\;\; \frac{1}{({\mu}-1)^2},
\end{equation}
where $F\equiv 16\pi^3n_0(c/H_0)^3(\sigma_v/c)^4$,
$n_0$ is the present comoving number density of the galaxies
assumed to develop as $n=(1+z_d)^3n_0$,
$\tilde{D}_d$, $\tilde{D}_s$ and $\tilde{D}_{ds}$ are the angular
diameter distances in units of $(c/H_0)$, corresponding to $D_d$, $D_s$ and
$D_{ds}$, respectively. In a way similar to the lensing probability by
pointlike masses, eq.(79) can be separated into two parts
by utilizing the optical depth $\tau$
%80
\begin{equation}
dp=d\tau\;\frac{1}{({\mu}-1)^2}.
\end{equation}
Figure 20 shows the differential optical depth $d\tau$ to
gravitational lensing for a source at different redshifts  $z_s=1,2,3$.
The significance of the differential optical depth is that it
provides a clear view of the most probable lens position for various sources.
It then turns out that for  quasars at their typical position of
$z_s\approx2$ the lensing galaxies locate most likely at $z_d\approx 0.5$.
%@@@@@@@@@@@@@@@@@@@@@@@@@@@@@@@@@@@@@@@@@@@@@@@@@@@@@@@@@@@@@@@@@@@@@@
%                          FIGURE 20
\begin{figure}
 \vspace{1cm}
 \caption{The differential optical depth to gravitational lensing by
          an ensemble of SIS galaxies. The cosmological model is
          chosen to be $\Omega=1$ and $\tilde{\alpha}=1$.
          $F\equiv 16\pi^3n_0(c/H_0)^3(\sigma_v/c)^4$. (cf. Turner,
          Ostriker and Gott, 1984)}
\end{figure}
%@@@@@@@@@@@@@@@@@@@@@@@@@@@@@@@@@@@@@@@@@@@@@@@@@@@@@@@@@@@@@@@@@@@@@@
The total optical depth $\tau$ can be obtained analytically
for some special cosmological models (Turner, Ostriker and Gott, 1984).
For example, in the case of $\Omega=1$ and $\tilde{\alpha}=1$ the total
optical depth is simply
%81
\begin{equation}
\tau=\frac{4F}{15}\frac{[(1+z_s)^{1/2}-1]^3}{(1+z_s)^{3/2}}.
\end{equation}
For a quasar at $z_s=2$ this reads
%82
\begin{equation}
\tau=2\times10^{-3}(F/0.1)
\end{equation}
An extensive analysis of
various types of galaxies by Fukugita and Turner (1991) gives
%83
\begin{equation}
\begin{array}{ll}
F=0.019\pm0.008 & \;\;\;\;\;{\rm for \;\; E\;\; galaxies}\\
F=0.021\pm0.009 & \;\;\;\;\;{\rm for \;\; S0 \;\; galaxies}\\
F=0.007\pm0.003 & \;\;\;\;\;{\rm for \;\; S \;\; galaxies}
\end{array}
\end{equation}
based on a morphological composition E:S0:S=12:19:69. Thus, for all the
galaxies as lenses, $F=0.047$. This results in a total optical depth of $0.001$
for a quasar at $z_s=2$, i.e., about $1/1000$ quasars at $z_s\sim2$ would be
found to be significantly lensed by foreground galaxies. Note that the
definition of the optical depth utilizes a cross-section of
$\pi\theta_E^2$, so that the total optical depth
in SIS corresponds to the total
probability of a background source being magnified by a factor of $\mu\geq2$.
Using other models (ISC, KING, $r^{1/4}$, etc.) for the matter distribution
of galaxies has yielded  statistical properties which
explain very well the observed frequency of lensed quasars,  the observed
distributions of image separations and of apparent magnitudes (Dyer, 1984;
Hinshaw and Krauss, 1987; Kochanek and Blandford, 1987; Wu, 1989b;
Mao, 1991; Fukugita and Turner, 1991; Kochanek, 1993a,b,c;1995; etc.).

It is worth noticing that the Hubble Space Telescope (HST) Snapshot Survey
(Bahcall et al., 1992; Maoz et al., 1992; Maoz et al., 1993a,b; Falco,
1993) provides a sample of 502 luminous and high redshift quasars, among
which a search for gravitationally lensed events has been made using the HST
Planetary Camera. This sample is of great significance for testing the
theoretical model of gravitational lensing which predicts that there should
be many gravitationally lensed quasars having $\sim 1$ arcsecond (Turner,
Ostriker and Gott, 1984) and even sub-arcsecond image separations (Fukugita
and Turner, 1991),
whilst the HST Snapshot Survey is capable of observing these small separated
cases.  One new candidate Q1208+1011 was found in the HST Snapshot Survey.
Together with the previously known cases in the sample, the observed
frequency of lensing is estimated to be between 3 and 6 out of 502 quasars.
The theoretically expected frequency of lensing can be obtained by
$\tau(z_s)B(m,z)$, where $B(m,z)$ is the factor by which lensed quasars
are over-represented among quasars of a given magnitude $m$ because fainter
quasars have been magnified to that magnitude, and its value is determined
by the quasar luminosity function (Bahcall et al. 1992).  As a consequence,
this indeed results in  a frequency compatible with the result of the HST
Snapshot Survey but does not meet the prediction by a cosmological
constant dominated Universe (Maoz et al., 1993b). The newly completed
lens quasar surveys have even limited $\lambda_0\equiv \Lambda/3H_0^2$ to
$\lambda_0<0.66$ (Kochanek, 1995).\\

%-------------------------------------------------------------------
\subsection{Determination of $H_0$}

The present status of the uncertainty of determination of the Hubble constant
$H_0$ by a factor of about two between $40$ km/s/Mpc and $80$ km/s/Mpc
is unfortunate, which has caused the cosmic distance dispute for decades.
The main problem arises from the disagreement of the ``standard candles" used
as the indicator of the absolute distance (see Fukugita, Hogan and Peebles,
1993 for a recent review).
Indeed, it is very unlikely that the debate on $H_0$  would be settled
in the next few years if the measurements are still based on the
conventional ``standard candle" methods. However, the recent progress of
determination of $H_0$ using other techniques that are independent of
the usual distance-ladder arguments may hopefully help to settle down
the debate. The time delay of the multiply-imaged quasars is one of
these methods which are promising  for the measurement of $H_0$.

The time that the photon traverses a proper distance of $\Delta r_{\rm prop}$
in the Universe is simply
%84
\begin{equation}
\Delta t=\frac{\Delta r_{\rm prop}}{c}=
\frac{1}{H_0}\frac{\Delta z}{(1+z)^2\sqrt{1+\Omega z}}\;\sim\;H_0^{-1}.
\end{equation}
Therefore, we can obtain the Hubble constant by measuring the time
difference $\Delta t$ for a given $\Omega$. The different
optical paths to earth between the multiple images of a lensed quasar
offer then a possibility of observing $\Delta t$ (the time delay) if
the quasar has intrinsic luminosity variability.

Actually, even before the discovery of the gravitationally-doubled quasars
it was realized that the Hubble constant $H_0$ could be measured from the
time delay of two images of a single background source (Refsdal, 1964a,b,
1966). The expression eq.(84) can now be generally written as
%85
\begin{equation}
H_0\Delta t\;=\;T(\Omega,z_d,z_s)\;
                f_{\rm lens}(\mbox{\boldmath $\theta_{A,B},\alpha$})
\end{equation}
where $T$ is called the cosmological correction function which is only
dependent on the cosmological parameters, and $f_{\rm lens}$ is the
lens model function which is given by the matter distribution of
the lens. $\mbox{\boldmath $\theta$}$ and $\mbox{\boldmath $\alpha$}$
are the positions of the images and
the deflection of light, respectively. It has been shown (Kayser, 1986)
that such a separation of the cosmological function $T$ from the lens model
function $f$ is indeed possible.

There are essentially three approaches developed to compute
the time delay, namely, the wavefront method (Kayser and Refsdal, 1983);
(2)the integration method (Cooke and Kantowski, 1975) and (3)the scalar
formulation (Schneider, 1985).
In the approximation of weak gravitational field ($\phi$), the traveling time
$t$ of a photon is (Cooke and Kantowski, 1975; Borgeest, 1983)
%86
\begin{equation}
ct\;=\;\int ds \;-\;\frac{2}{c^2}\int\phi ds.
\end{equation}
The first term gives the length of the light path, and the second is the
relativistic time dilatation due to the gravitational potential $\phi$ of
the deflector. The integrals are performed along the photon orbit $ds$.
Correspondingly, the propagation delay for light-rays from the double images
of a lensed source can be split into two components: (1)the geometrical delay
$\Delta t_g$ that is caused by the difference of light paths between images
and (2)the potential delay $\Delta t_p$ that is induced by the difference of
the gravitational field of the intervening lensing object
at the image positions.
A straightforward computation gives the two components
(Cooke and Kantowski, 1975; Borgeest, 1983; Borgeest and Refsdal, 1984):
%87,88
\begin{eqnarray}
c\Delta t_g=(1+z_d)D_d \frac{\mbox{\boldmath $\alpha(\theta_1)$}+
\mbox{\boldmath $\alpha(\theta_2)$}}{2}\cdot
(\mbox{\boldmath $\theta_1$}- \mbox{\boldmath $\theta_2$})\\
c\Delta t_p=(1+z_d)D_d \int_{\mbox{\boldmath $\theta_1$}}
^{\mbox{\boldmath $\theta_2$}}\mbox{\boldmath $\alpha$}
(\mbox{\boldmath $\theta$})\cdot d \mbox{\boldmath $\theta$}.
\end{eqnarray}
Finally, extracting $H_0$ from $D_d$ and combining it with
$\Delta t=\Delta t_g+\Delta t_p$ yields the form of eq.(85).

A detailed lens model has been developed for the well-known lens system
QSO0957+561 in order to determine the Hubble constant $H_0$ from the
gradually accumulated data of the light curves of the double
quasar images A,B (Young et al., 1980;
Dyer and Roeder, 1980; Greenfield, Roberts and Burke, 1985; Falco, Gorenstein
and Shapiro, 1985;1991; Gorenstein, Falco and Shapiro, 1988;  etc.). A
lensing model that is composed of three matter components has been found
to reproduce the known properties of QSO0957+561A,B quite well: (1)the bright
galaxy (G1) described by a King profile ($\sigma_v$, $r_c$);
(2)a compact nucleus with mass of $M_c\sim10^{11}h^{-1}M_{\odot}$ and
(3)the cluster characterized by the surface mass density of a smoothly
distributed mass screen (Gorenstein, Falco and Shapiro, 1988; Falco,
Gorenstein and Shapiro, 1991).  This lens model leads to a relatively
simple expression of the Hubble constant $H_0$ from eqs.(85), (87) and (88)
(see also Roberts et al., 1991)
%89
\begin{equation}
H_0=\left\{
\begin{array}{l}
97\pm20\\
90\pm21
\end{array} \right\}
\left(\frac{\sigma_v}{390\;{\rm km/s}}\right)^2\;
\left(\frac{1\;{\rm yr}}{\Delta t}\right)\;{\rm km/s/Mpc}\;\;
\left\{
\begin{array}{l}
\Omega=0\\
\Omega=1
\end{array}\right\}.
\end{equation}
The error estimate includes measurement error between the VLBI, VLA and
optical observations,  the unknown values of
$\Omega$ and of the clumpiness of the Universe, the non-uniqueness of
the cluster model, and errors in the detailed model of $G1$.
The result depends weakly on the cosmological density
parameter $\Omega$.  It appears that two free parameters, $\sigma_v$ and
$\Delta t$, control the actual evaluation of $H_0$.  Motivated by the
significance of determination of $H_0$ from the time delay, Rhee (1991)
obtained the line-of-sight velocity dispersion of $303\pm50$ km/s for the
bright galaxy G1 in the QSO0957+561 lens system, leaving the final work of
finding $H_0$ to the measurement of the time delay between the double images.
However, it should be mentioned that the simple lens model from which
eq.(89) was derived is by no means unique, and more complicated mass
distributions are not only possible but also actually well motivated
(Bernstein, Tyson and Kochanek, 1993).

Vanderriest et al. (1989) undertook a 8-years optical photometric
monitoring of
the double quasar 0957+561A,B from 1980 to 1987, which contains totally
131 observations. The light curves of the image A and B
are shown in Figure 21.
%@@@@@@@@@@@@@@@@@@@@@@@@@@@@@@@@@@@@@@@@@@@@@@@@@@@@@@@@@@@@@@@@@@@@@@
%                          FIGURE 21
\begin{figure}
 \vspace{1cm}
 \caption{Optical light curves of the double images of QSO0957+561.
          (From Vanderriest et al., 1989)}
\end{figure}
%@@@@@@@@@@@@@@@@@@@@@@@@@@@@@@@@@@@@@@@@@@@@@@@@@@@@@@@@@@@@@@@@@@@@@@
A significant decrease in brightness around Julian Days 2445700 in A and
Julian Days 2446100 in B is clearly seen, indicative of a time delay $\sim
400$ days. The original analysis of Vanderriest et al. (1989) from the
cross-correlation function for the two light curves gives $\Delta t=415$ days,
while Press, Rybicki and Hewitt (1992) reached a value of
$\Delta t=(537\pm11)$ days based on a newly
developed mathematical methodology
for the same data.  Using additional optical data of 3.5 years coverage,
Schild (1990) obtained a value of $\Delta t=404$ days, consistent with
the result of Vanderriest et al. (1990). The updated optical light curves to
1994 July (Figure 22) seems also to support $\Delta t\approx 1.1$ years
(Schild and Thomson, 1995).
%@@@@@@@@@@@@@@@@@@@@@@@@@@@@@@@@@@@@@@@@@@@@@@@@@@@@@@@@@@@@@@@@@@@@@@
%                          FIGURE 22
\begin{figure}
 \vspace{1cm}
 \caption{The updated optical light curves of the double images of
	  QSO0957+561 A (upper) and B (lower) with the B data retarded by
	  1.1 years and offset by 0.2 mag. The 832 actual observations
	  of the image A are shown at the bottom.
          (From Schild and Thomson, 1995)}
\end{figure}
%@@@@@@@@@@@@@@@@@@@@@@@@@@@@@@@@@@@@@@@@@@@@@@@@@@@@@@@@@@@@@@@@@@@@@@
Radio monitoring of QSO0957+561A,B
with the VLA was reported later for a 10 -- 11 years coverage (Roberts et al.,
1991; Leh\'ar et al., 1992).  The flux curves of the two images are shown
in Figure 23. In the absence of specific features in the two curves,
%@@@@@@@@@@@@@@@@@@@@@@@@@@@@@@@@@@@@@@@@@@@@@@@@@@@@@@@@@@@@@@@@@@@@@@
%                          FIGURE 23
\begin{figure}
 \vspace{1cm}
 \caption{Radio flux curves of the A (uppercase) and B (lowercase)
          images of the double QSO0957+561. The letters denote the VLA
          configurations (P for partial configurations).
          (From Leh\'ar  et al., 1992)}
\end{figure}
%@@@@@@@@@@@@@@@@@@@@@@@@@@@@@@@@@@@@@@@@@@@@@@@@@@@@@@@@@@@@@@@@@@@@@@
an analysis of the cross-correlation function of two signals suggests
a time delay of $513\pm40$ days. The very recent measurement of the
time delay using the hybrid maps of QSO0957+561A,B with
VLBI spanning the 6-year interval (1987--1993) yields
$\Delta t\sim 1$ year (Campbell et al., 1995), while a reanalysis of
the Leh\'ar et al. (1992) observation with a refined method has found that
their radio data are compatible with the result of $\sim1$ year obtained
from the optical data (Pelt et al. 1995). Moreover,
the updated VLA light curves of QSO0957+561A,B for 16 years
coverage show a time delay of $\Delta t=455\pm40$ days
(Haarsma et al., 1995). Apparently, the
present results of measurement of the time delay in the double
images QSO0957+561A,B are controversial,
and the acceptable value  ranges from 404 days to 537 days.

Adopting the cosmological density parameter of $\Omega=1$ and the
velocity dispersion of $\sigma_v=303\pm50$ km/s for the G1 in
QSO0957+561A,B system, one finds from eq.(89)
%90
\begin{equation}
H_0=\left\{
\begin{array}{l}
48^{+16}_{-7}\\
39^{+13}_{-6}
\end{array}\right\}
{\rm km/s/Mpc}\;\;\left\{
\begin{array}{l}
\Delta t=415\pm20\;{\rm days}\;\;{\rm (optical)}\\
\Delta t=513\pm40\;{\rm days}\;\;{\rm (radio)}
\end{array}\right\}.
\end{equation}
Though these values still contain large uncertainties, they seem to support a
low value of the Hubble constant. Further observations will be needed
to find a reliable value of time delay in QSO0957+561A,B as well as
to establish a reliable lensing model
in order to precisely determine the value of $H_0$.
The new observation of luminosity variations in the
quadruple-lens system B1422+231 (Hjorth et al, 1995)
and the recent detection of time delay in the Einstein
ring B 0218+367  (Corbett et al., 1995) and
PKS 1830-211 (van Ommen and Preston, 1995) would be also promising
for the measurement of $H_0$.\\

%-------------------------------------------------------------------
\subsection{Quasar-galaxy associations}

One of the important consequences of gravitational lensing, as first realized
by Gott and Gunn (1974) even before the discovery of the first
gravitationally-imaged quasar 0957+561A,B,
is that the surface number density of quasars near
foreground galaxies would be enhanced (denoted by the quasar enhancement factor
$q_Q$) because the distant quasars lying behind galaxies would be magnified
by the lensing effect of the galaxies and then enter into the detection limit
(see also Canizares, 1981; Vietri and Ostriker, 1983; Schneider,
1986;1987a,b; Kovner, 1989; etc.).
Equivalently, an overdensity of foreground galaxies
around high-redshift quasars would also exist (described by the galaxy
enhancement factor $q_G$) (Schneider, 1989). The statistical evidence on such
quasar-galaxy associations was firstly found by Tyson (1986) and later
reported by Webster et al. (1988). They all claimed a significant enhancement
of galaxy surface density in the vicinity of distant quasars.
Since then, the observational evidences for quasar-galaxy associations have
been cumulated (Narayan, 1992). Table 4 summarizes the present status on
the optically-selected quasar-galaxy associations,
including two negative results.  Some suggestions
have been made to improve the confidence of the different results
such as choosing the same objects, cross-calibrating the different observing
techniques, using the same criteria, etc.
Yet, large samples will be needed to
further confirm the existence of quasar-galaxy associations.
%*****************************Table 4 **************************************
   \begin{table}
      \caption{Foreground galaxy enhancement $q_G$}
         \label{ }
      \[
         \begin{array}{c|c|c|c|c|c}
            \hline
            {\rm authors}  & {\rm QSO} & {\rm selections}
			   & \theta\; {\rm range} ('') &
                             {\rm galaxy} (R)  &q_G\\
            \hline
	   {\rm Crampton} & 101 & V<18.5 & 0-6 & \sim23& 1.4\pm0.5\\
	                  &     & z>1.5  &     &        &  \\
	    \hline
	   {\rm Kedziora-} & 181 & V<18.5 & 6-90 & \sim21.5&\sim 1\\
	   {\rm Chudczer} &     & z>0.65 &    &     & \\
	    \hline
           {\rm Magain} & 153 & \overline{V}=17.4 & 0-3 & \sim21&\sim2.8\\
                        &     & \langle z\rangle=2.3 &   &  & \\
	    \hline
	   {\rm Thomas} &    64  & V<18.5 & 0-10 & \sim22& \sim1.7 \\
               &        &     1<z<2.5       &  &  & \\
	    \hline
	   {\rm Van\; Drom} & 136 & \overline{V}=17.4 & 3-13.7 &
						 \sim23& \sim1.46 \\
                          &     & \langle z\rangle=2.3 & &    &      \\
	    \hline
	   {\rm Webster} & 68 & V<18 & 3-10  &\sim22& \sim2\\
                         &    & 0.7\!\!<\!\!z\!\!<\!\!2.3 &   & & \\
	    \hline
	   {\rm Yee} &    94  & V<19 & 2-6 & \sim22.5& 1.0\pm0.3 \\
               &          &     z>1.5       & 2-10 &  &1.0\pm0.2\\
               &        &                 & 2-15 & &0.9\pm0.1\\
	    \hline
         \end{array}
      \]
   \end{table}
%************************************************************************

The first effect of gravitational lensing is its magnification (${\mu}$),
which enhances the apparent brightness of background sources by an amount of
$2.5\log {\mu}$ in magnitude, leading to an increase of the
surface number density ($\sigma$) of
background sources by picking up the faint sources:
$\sigma(\theta)\sim N(<m+2.5\log {\mu})/S_0(\theta)$,
where $S_0(\theta)$ is the
observed area at a distance $\theta$ from the deflector.
The second effect is the area distortion $S_0(\theta)+\Delta S_0(\theta)$,
which arises from the light
bending around the deflector (see Figure 24). This reduces the number counts
%@@@@@@@@@@@@@@@@@@@@@@@@@@@@@@@@@@@@@@@@@@@@@@@@@@@@@@@@@@@@@@@@@@@@@@
%                          FIGURE 24
\begin{figure}
 \vspace{1cm}
 \caption{Scheme of two effects of gravitational lensing on the number counts.
          Magnification effect enhances the apparent magnitude
	  of the background sources, which helps to pick up the fainter
	  sources  and then leads to an increase of the total
          number of the sources in a flux-limited sample. The area distortion
          effect due to the light deflection reduces the source volume
          (the dashed-line region), resulting in a decrease of the total
          number counts.}
\end{figure}
%@@@@@@@@@@@@@@@@@@@@@@@@@@@@@@@@@@@@@@@@@@@@@@@@@@@@@@@@@@@@@@@@@@@@@@
by losing the sources within the dashed-line regions:
$\sigma(\theta)\sim N(<m)/(S_0(\theta)+\Delta S_0(\theta))$.
As a whole, the surface number density can be written as
%91
\begin{equation}
\sigma(\theta)=\frac{N(<m+2.5\log {\mu})}{S_0(\theta)+\Delta S_0(\theta)}.
\end{equation}
Defining the enhancement factor $q_Q(\theta)$ as the ratio of the disturbed
surface number density $\sigma(\theta)$ to the undisturbed one
$\sigma_0(\theta)=N(<m)/S_0(\theta)$ and noticing that
$\mu(\theta)=[S_0(\theta)+\Delta S_0(\theta)]/S_0(\theta)$, one has
(Narayan, 1989)
%92
\begin{equation}
q_{Q}(\theta)=\frac{N(<m+2.5\log \mu(\theta))}{N(<m)}\;\frac{1}{\mu(\theta)}.
\end{equation}
It appears that the ``local" enhancement parameter $q_{Q}$ at
$\theta$  depends on two factors: the intrinsic
number-magnitude relation of the background sources and the local
lensing magnification around $\theta$.  The average quasar enhancement
$\bar{q}_Q(\theta_1,\theta_2)$ over an angular distance
of $(\theta_1,\theta_2)$
from the foreground deflector is simply
%93
\begin{equation}
\bar{q}_{Q}(\theta_1,\theta_2)=\frac{2\int_{\theta_1}^{\theta_2}
q_{Q}(\theta)\;\theta\;d\theta}{(\theta_2^2-\theta_1^2)}.
\end{equation}

Two kinds of quasar number-magnitude relations have been thus far suggested
from, respectively, the BSP survey and the HV survey.
The significant difference
in these two relations is that the slope of the BSP counts changes at $B\approx
19.15$ from $0.86$ to $0.28$ while there is no such a turnover in the range of
$B\leq21$ in the HV counts (see Figure 15). This probably arises from the
different selection methods used in the two surveys.
The BSP cumulative counts can be fitted by (Narayan, 1989)
%94
\begin{equation}
\begin{array}{ll}
N(<B)=4.66\times10^{0.86(B-19.15)}, & B<19.15;\\
N(<B)=-10.95+15.61\times10^{0.28(B-19.15)}; \;\; & B>19.15.\\
\end{array}
\end{equation}
This relation is valid for $z\leq2.2$ and $B<21$. Nevertheless, the subsequent
observation (Boyle, Jones and Shanks, 1991) indicates that the above relation
holds true also to $B\leq22$. The HV cumulative counts can be fitted by
(Wu, 1994a)
%95
\begin{equation}
N(<B)=6.25\times10^{0.51(B-19.15)}.
\end{equation}

The local enhancements $q_{Q}$ are computed for these two kinds of quasar
number counts and plotted against the local magnification ${\mu}$
in Figure 25. One should pay a special attention to the case where
$B+2.5\log {\mu}$ is larger than the limit of validity of the
quasar number count relation.
BSP and HV surveys were both restricted within $B\leq21$. Therefore, the
number-magnitude relation $N(<B+2.5\log {\mu})$
fails when $B+2.5\log {\mu}>21$, which occurs for a sufficiently
large ${\mu}$. Strictly speaking, one cannot calculate the
%@@@@@@@@@@@@@@@@@@@@@@@@@@@@@@@@@@@@@@@@@@@@@@@@@@@@@@@@@@@@@@@@@@@@@@
%                          FIGURE 25
\begin{figure}
 \vspace{1cm}
 \caption{The local quasar enhancement $q_Q$ against quasar limiting magnitude
          $B$ and local lensing magnification $\mu$ for the BSP
          and the HV counts.
          The solid lines correspond to the results within the BSP and HV
	  survey limit of $B=21$, i.e., both $B$ and $B+2.5\log\mu$ in
	  the computation of $q_Q$ are confined to  $B<21$,
	  and the dotted lines are the extrapolated results by employing
          $N(<B)$ beyond $B=21$. Note that in the HV counts, $q_Q\geq1$,
          providing always the positive associations, while in the BSP counts
          $q_Q$ may be smaller than unity, leading to  the ``negative"
          associations.}
\end{figure}
%@@@@@@@@@@@@@@@@@@@@@@@@@@@@@@@@@@@@@@@@@@@@@@@@@@@@@@@@@@@@@@@@@@@@@@
enhancement $q_Q$ beyond the survey limit, and the extrapolation of the solid
lines in Figure 25 requires the knowledge of fainter quasar counts.
Furthermore, the application of eqs.(94) and (95) for the evaluation of
$q_Q$ in eq.(92) has presumed that the observed $N(<B)$ remains the same as
the intrinsic counts, i.e., quasar counts
have not been contaminated significantly by lensing.

A power-law number-magnitude relation with index of $\alpha$, $N(<m)\sim
10^{\alpha m}$, would lead to $q_Q=\mu^{2.5\alpha-1}$, independent of the
limiting magnitude $m$. The HV counts have $\log q_Q/\log {\mu}=0.3$,
which then cannot provide a large enhancement for a moderate magnification.
Conversely,
the two power-laws of BSP counts give rise to a relatively wide range of
$q_Q$, depending on both the limiting magnitude and the magnification.
Bright quasars ($B<18$) appear to be relatively strongly associated with
foreground deflectors, with a maximum enhancement at $dq_Q/d{\mu}=0$.
On the other hand, faint quasars ($B>19.15$) exhibit a ``negative"
association with the foreground deflectors, i.e., fewer quasars would be
found near the foreground galaxies than in the rest of the sky.

Adopting a SIS model for the matter distribution of a lensing galaxy, we can
express the local magnification at an angular distance $\theta$ from the
center of the galaxy as  [eq.(27)]
%96
\begin{equation}
\mu=\frac{\theta}{\theta-\theta_E}.
\end{equation}
Furthermore,  we assume the distance parameter $D_{ds}/D_s$ to be very close
to unity, which approximately holds true when the foreground galaxies are at
relatively low redshift while quasars are at high redshift in order to
guarantee that they are not physically associated systems in the searches for
quasar-galaxy associations.  In this case, the Einstein radius reads
%97
\begin{equation}
\theta_E=1''.33\left(\frac{\sigma_v}{215{\rm km/s}}\right)^2.
\end{equation}
Figure 26 shows the average enhancement $\bar{q}_Q$ versus
the limiting magnitudes
and the search ranges around a typical galaxy of $\sigma=215$ km/s, the
average of the E/S0 galaxy velocity dispersions (Kochanek, 1993c),
provided that the
extrapolation of both BSP and HV counts to the faint magnitude ($B>21$)
is possible. In HV counts $\bar{q}_Q$ depends only on the search areas and
the resulting amplitude turns to be too small to  explain
the reported enhancements of as large as 2 listed in Table 4.
Both positive and ``negative" associations are provided by BSP counts,
separated in the range of $19<B<20$. Other important conclusions are:
 (1)Positive associations between foreground
galaxies and background quasars would be found when one chooses the
limiting magnitude
of the quasar sample to be brighter than $B\approx19$. (2)When the
faint quasars ($B>19.5$) are involved, one would expect to detect null
or ``negative" associations.
%@@@@@@@@@@@@@@@@@@@@@@@@@@@@@@@@@@@@@@@@@@@@@@@@@@@@@@@@@@@@@@@@@@@@@@
%                          FIGURE 26
\begin{figure}
 \vspace{1cm}
 \caption{The average quasar enhancement $\bar{q}_Q$ over different
	  search areas around a foreground galaxy with $\sigma_v=215$ km/s.}
\end{figure}
%@@@@@@@@@@@@@@@@@@@@@@@@@@@@@@@@@@@@@@@@@@@@@@@@@@@@@@@@@@@@@@@@@@@@@@
This scenario of existence of positive/negative quasar-galaxy associations
can explain the observed results (Table 4) quite well (Wu, 1994a).
In fact, the search for  ``negative" associations between quasars and
galaxies at faint magnitude can be used as the ultimate test for whether
or not the quasar-galaxy associations stem from the effect of gravitational
lensing.   There are no other mechanism known thus far that can result in
the negative associations, i.e., the surface number density of distant
quasars around foreground galaxies is smaller than the mean quasar
density in the rest of the sky.

For foreground galaxies with different luminosities which follow
the Schechter function
$\phi(L)dL=\phi^*(L/L_*)^{\nu}$ $\exp(-L/L_*)d(L/L_*)$,
we can find the average quasar enhancement over a search range of
$(\theta_1,\theta_2)$ from
%98
\begin{equation}
\frac{\langle qN\rangle}{\langle N\rangle}=
\frac{\displaystyle\sum_i
\int_{L_{min,i}}^{\infty}\bar{q}_{Q,i}(\theta_1,\theta_2)
     \gamma_i\phi_i(L)dL}
    {\displaystyle\sum_i\int_{L_{min,i}}^{\infty}\;
	\gamma_i\phi_i(L)dL},
\end{equation}
where $i$ and $\gamma_i$ represent, respectively, the $i$-th type and
composition of galaxies, and the minimum galaxy luminosity
$L_{min}$ is related to the galaxy limiting
magnitude in the observation of quasar-galaxy associations.
The galaxy luminosity $L$ can be converted
into the velocity dispersion $\sigma_v$ through the empirical
formula such as the Faber-Jackson relation for early-type galaxies
(E/S0) $L/L_*=(\sigma_v/\sigma_*)^4$ or the Tully-Fisher relation for
spiral galaxies (S) $L/L_*=(\sigma_v/\sigma_*)^{2.6}$. The parameters
$L*$ (or $\sigma_*$), $\phi^*$ and $\nu$ have been observationally
determined for different galaxies (see Fukugita and Turner, 1991).
Finally, the spatial distribution of galaxies needs to be
taken into account.
Assuming a constant comoving number density of galaxies, we have
%99
\begin{equation}
\langle \bar{q}_Q(\theta_1,\theta_2)\rangle=
\frac{\int_0^{z_s}4\pi D_d^2 (1+z_d)^3\;
	\langle qN\rangle\;dr_{prop,z_d}}
     {\int_0^{z_s}4\pi D_d^2 (1+z_d)^3\;
        \langle N \rangle\;dr_{prop,z_d}},
\end{equation}
in which $dr_{prop,z_d}=(c/H_0)dz_d/[(1+z_d)^2\sqrt{1+\Omega z_d}]$.
This theoretical expectation can be used straightforwardly to
compare with the observations (Wu, Zhu \& Fang, 1995).

It should be realized that the quasar-galaxy associations are the
statistical results and thereby, statistical lensing should be
involved. An exact treatment of this question is to convolve
the magnification probability $P(\mu)$ by foreground galaxies
with the intrinsic quasar number counts $N(<m)$ (Schneider, 1989).
The difficulty is that $P(\mu)$ should be artificially truncated
at the faint end of magnification in order to perform
the integration $\int\;N(<m)dP(\mu)$.

\bigskip
%@@@@@@@@@@@@@@@@@@@@@@@@@@@@@@@@@@@@@@@@@@@@@@@@@@@@@@@@@@@@@@@@@@@@@@@
\bigskip
\section{CLUSTERS OF GALAXIES AND
ARCLIKE IMAGES}

\subsection{Giant arcs and arclets}

The arclike image associated with Abell cluster 370 was first detected by
Hoag in 1981. However, this blue arc had not been recognized to be
the image of a distant galaxy gravitationally lensed by the cluster until
a few years later when two groups of astronomers independently announced
their convincing evidences of the existence of this peculiar feature in
the Universe (Soucail et al., 1987a; Lynds and Petrosian, 1986), followed
by Paczy\'nski's (1987) lensing interpretation.
Only two giant arcs (A370 and Cl2244-02, see Figure 27)
were known at that time and other explanations remained also  possible.
The crucial point of interpretation of the arcs as  gravitationally
imaged background sources rather than some peculiar features physically
associated with the clusters is the measurements of redshift of the
two giant arcs.  They do show higher redshifts than those of their
associated clusters: The giant arc in Abell 370 has a redshift of 0.725
(Soucail et al., 1987b; Miller and Goodrich, 1988), in comparison
%@@@@@@@@@@@@@@@@@@@@@@@@@@@@@@@@@@@@@@@@@@@@@@@@@@@@@@@@@@@@@@@@@@@@@@
%                          FIGURE 27
\begin{figure}
 \vspace{1cm}
 \caption{Giant luminous arcs in Abell 370 (a) (Figure courtesy of G. Soucail)
          and Cl 2244-02 (b) (Figure courtesy of F. Hammer)}
\end{figure}
%@@@@@@@@@@@@@@@@@@@@@@@@@@@@@@@@@@@@@@@@@@@@@@@@@@@@@@@@@@@@@@@@@@@@@@
with the redshift of 0.374 for Abell 370 itself , and the arc in Cl 2244-02
has probably an even higher redshift of 2.237 (Mellier et al., 1991)
while its associated cluster is only at $z_d=0.336$.  These measurements have
strongly confirmed the lensing origin of arclike images seen in the cores
of rich galaxy clusters. Up to now giant arcs and arclets have been detected
in about 30 clusters of galaxies (Table 5) and this number is still
increasing dramatically.
%**************************Table 5 ****************************************
   \begin{table}
      \caption{Arclike Images}
           \label{ }
      \vspace{1cm}
   \end{table}
%**************************************************************************
In particular, the high X-ray luminosity clusters in the EMSS sample
turn to be the very efficient deflectors of producing  arcs
and about 14 arcs have thus far been seen in a subsample of 41 EMSS clusters
with $L_x\geq2\times10^{44}{\rm erg/s}$ and $z_d\geq0.15$
(Hammer et al., 1993; Le F\`evre et al., 1994; Gioia and Luppino, 1994;
Hammer, 1995; Luppino et al., 1995).
Note that there are several multiple arc systems.

Wu and Hammer (1993) classified the elongated images
using two parameters: axial ratio ($L/W$, i.e., length/width) and apparent
magnitude ($B$). The ``giant" arcs refer to those images whose axial ratios
are greater than 10, i.e., $L/W\geq10$,  and mini-arcs or arclets
have $L/W\leq3$. The rest arclike images in between ($3<L/W<10$) are called
medium arcs. Arc brightness is represented by its $B$ magnitude
so that the ``luminous" arcs have $B\leq22.5$. In the updated list of
arclike images of Table 5 there are totally  $\sim10$ giant luminous arcs,
while arclets are numerous but usually very faint ($B\sim26$).  \\

%-------------------------------------------------------------------------
\subsection{Clusters as lenses}

Although most of the arclike images remain to be spatially unresolved
in width today,
the colours and spectra of the arcs are compatible with
those of local sub-$L_*$ spiral galaxies, indicating that they are probable
spirals at relatively high redshift $z_s\sim1$.  The foreground clusters
often have high X-ray luminosity ($L_x>10^{44}$ erg/s) and/or large velocity
dispersion ($\sigma_v>1000$ km/s), which are then massive
enough to act as strong lenses for the background galaxies.

To demonstrate how a rich galaxy cluster at intermediate redshift
gravitationally distorts background galaxies, a simulation is made based
on the current knowledge of dynamical and spatial properties of galaxies and
rich clusters of galaxies. SIS is adopted for the matter distribution
of a galaxy cluster with $\sigma_v=1000$ km/s at $z_d=0.25$. The luminosity
function established  by Broadhurst, Ellis and Shanks (1989) in deep redshift
survey is employed for the distribution of background spiral galaxies. The
luminous area of each galaxy is taken to be a circular
disk of radius of $R$ which
is assumed to follow the relation $R=R_*(L/L_*)^{1/2}$, where $R_*$ is the
characteristic radius corresponding to a $L_*$-galaxy.
In the actual simulation, $R=10^{(-17.16-M)/5}$ kpc (Freeman, 1970).
Figure 28(a) shows the unperturbed
%@@@@@@@@@@@@@@@@@@@@@@@@@@@@@@@@@@@@@@@@@@@@@@@@@@@@@@@@@@@@@@@@@@@@@@
%                          FIGURE 28
\begin{figure}
 \vspace{1cm}
 \caption{(a)The projected galaxy distribution on the sky without lensing
	  and (b) the distorted and magnified images of background galaxies
	  by a galaxy cluster (at center) with $\sigma_v=1000$ km/s
          and at $z_d=0.25$.}
\end{figure}
%@@@@@@@@@@@@@@@@@@@@@@@@@@@@@@@@@@@@@@@@@@@@@@@@@@@@@@@@@@@@@@@@@@@@@@
background galaxies projected on the sky in a field of $1'\times1'$,
in which the orientations of the disk galaxies are randomly placed in
space and the population of the galaxies is presumed not to evolve
cosmologically. Moreover, we truncate the redshift of the galaxies at
$z_s=1.25$ due to the
failure of the $K$-correction in the adopted luminosity function
(Broadhurst, Ellis and Shanks, 1989). Figure 28(b)
illustrates  the same field with a
galaxy cluster at the center. All the images brighter than $B=23$ are
selected by taking the magnification effect into account.  It appears that
galaxies in the field have been strongly elongated around the cluster,
indicating that distant rich clusters of galaxies can
indeed act as strong lenses and produce the images of giant arcs and arclets.

An interesting issue is the total number of mini-arcs and medium arcs
with respect to the number of giant arcs seen up to the same flux threshold
in a galaxy cluster. A statistical distribution of arc number
appearing in Figure 28(b) against axial ratio
$L/W$ is plotted in Figure 29, in which the normalization is made at $L/W=10$.
%@@@@@@@@@@@@@@@@@@@@@@@@@@@@@@@@@@@@@@@@@@@@@@@@@@@@@@@@@@@@@@@@@@@@@@
%                          FIGURE 29
\begin{figure}
 \vspace{1cm}
 \caption{The relative distribution  (normalized at $L/W=10$) of
	  arcs predicted in Figure 28(b) against their axial ratio.}
\end{figure}
%@@@@@@@@@@@@@@@@@@@@@@@@@@@@@@@@@@@@@@@@@@@@@@@@@@@@@@@@@@@@@@@@@@@@@@
It turns out that for one observed giant arc, there
should be another one medium arc of $L/W\approx6$ and two arclets of
$L/W\approx3$. The more careful statistical investigations have
reached essentially similar conclusion (Grossman and Narayan, 1988;
Wu and Hammer, 1993).  Unfortunately, this theoretically
predicted axial ratio distribution
has not been seen in the recent arc survey with the subsample of EMSS
clusters. Among the 14 arcs found in the 41 EMSS clusters, 9 are giant
arcs (Hammer et al., 1995; Luppino et al., 1995).
It is believed that the asymmetrical matter distribution of the arc clusters
may account for the discrepancy (Bartelmann and Weiss, 1994; Bartelmann,
Steinmetz and Weiss, 1995).

Modeling the known giant luminous arcs, even arclets,
associated with clusters
%@@@@@@@@@@@@@@@@@@@@@@@@@@@@@@@@@@@@@@@@@@@@@@@@@@@@@@@@@@@@@@@@@@@@@@
%                          FIGURE 30
\begin{figure}
 \vspace{1cm}
 \caption{Modeling the giant luminous  arc and arclets in Abell 370.
          Upper left: The direct CCD image of the core of Abell 370;
          Upper right: Image construction. Solid lines and dashed lines
	  are, respectively, the caustics and critical lines (the inner
	  one is for $z_s=0.725$ and the outer one, $z_s=0.895$).
	  Dot-dashed lines are the
          core radius, ellipticity and orientation of the two matter
          profiles; Low left: Contours of the  surface density;
          Low right: Reconstruction of the source positions
	  (see the upper right panel for the corresponding part).
	  (from  Kneib et al., 1993)}
\end{figure}
%@@@@@@@@@@@@@@@@@@@@@@@@@@@@@@@@@@@@@@@@@@@@@@@@@@@@@@@@@@@@@@@@@@@@@@
of galaxies turns to be very successful. A simple elliptical or bimodal
potential for the arc cluster can provide the major properties compatible
with the observed arcs.  Indeed, any type of elongated configurations
including
the straight arcs and the radial arcs has been well reproduced based on
elongated gravitational potentials tracing the luminous matter distributions
(Pello et al., 1991; Mellier, Fort and Kneib 1993). An example of modeling
the arcs in the well-known Abell cluster 370 is shown in Figure 30.
This arc-cluster system has been extensively studied by a number of authors
(e.g. Hammer and Rigaut, 1989; Grossman and Narayan, 1989),
in particularly by the Toulouse Group
(e.g. Kneib et al., 1993; Soucail and Mellier, 1993).

%-------------------------------------------------------------------------
\subsection{Cluster matter distributions from arcs}

Arclike images are robust matter estimators of clusters
of galaxies, which is actually the most important issue of studying arclike
images today. From the general expression of
the lensing equation [eq.(12)] for
a spherical matter distribution [eq.(8)], we
can write out the projected mass along the line of sight within the arc
position $\theta$ ( in arcseconds) to be
%100
\begin{equation}
m_g(\theta)=7.37\times10^{11}(\theta-\beta)\theta
    \frac{\tilde{D}_d \tilde{D}_{ds}}{\tilde{D}_s}
       \;M_{\odot}\;h_{50}^{-1},
\end{equation}
where $\beta$ is the alignment parameter of background galaxy in arcseconds.
This parameter, however, is unmeasurable in practice.
Nevertheless, if we further assume a SIS model for
the matter distribution of cluster of galaxies and a uniform
circular disk
for background source,  the maximum width ($W$) of the arclike
image will be the same as the size of the background galaxy and
$\beta$ will satisfies the
following approximate geometrical relation
%101
\begin{equation}
\beta=\frac{L}{2(L/W)\sin(L/2\theta D_d)}.
\end{equation}
Therefore, the gravitational
mass contained in the central core ($<\theta$) of an arc cluster
is obtained without any assumptions about the dynamical state of the system.
For most of the arc-cluster systems, the gravitational mass is typically
$\sim10^{14}M_{\odot}$ in the core of a galaxy cluster.  The
significance of this method is that it provides an independent way of
calculating the masses of clusters of galaxies, which can be compared directly
with the masses estimated from the dynamical analysis based on the virial
theorem. Recall that the latter presumes an hydrostatic equilibrium  of both
the hot intracluster gas and the galaxies with the binding cluster potential
(Cowie, Henriksen and Mushotzky, 1987). This hypothesis,  unfortunately,
has not been verified using other astrophysical means.

Assuming that both the X-ray gas and the galaxies are in hydrostatic
equilibrium with the binding gravitational potential of a spherical galaxy
cluster, one can obtain the virial mass of the cluster
within radius of $r$ through
%102
\begin{equation}
M_v(r)=-\frac{kTr}{G\mu_p m_p}\left(\frac{d\ln n}{d\ln r}+\frac{d\ln T}{d\ln r}
\right),
\end{equation}
where $\mu_p$ is the mean particle weight in unit of the proton mass $m_p$,
$n$ and $T$ are the gas density and temperature, respectively, which can be
found by inverting the observed X-ray surface brightness profile, namely,
the $\beta$ model
%103
\begin{equation}
S(\theta)=S_0\left[1+(\theta/\theta_c)^2\right]^{1/2-3\tilde{\beta}}
\end{equation}
with a core radius of $\theta_c$ (or $r_c$ in linear size).
The isothermal gas distribution has been
found to be consistent with the X-ray observations of galaxy clusters,
leading to $d\ln T/d\ln r=0$. So,
the projected virial mass $m_v(\theta)$ from eq.(102) within a radius of
$\theta$ on the cluster plane is (Wu, 1994c)
%104
\begin{equation}
m_v(\theta)=1.14\times10^{14}\tilde{\beta}\tilde{m}(\theta)
	\left(\frac{kT}{\rm keV}
       \right) \left(\frac{r_c}{{\rm Mpc}}\right)\;M_{\odot};
\end{equation}
%105
\begin{equation}
\tilde{m}(\theta)=\frac{R_0^3}{1+R_0^2}-\int_{\theta/\theta_c}^{R_0}
      x\sqrt{x^2-\frac{\theta^2}{\theta_c^2}^2}
	\frac{3+x^2}{(1+x^2)^2}dx.
\end{equation}
Here $R_0=R/r_c$, and $R$ is the physical size of the cluster.
The numerical computations show that
$m(\theta)$ remains nearly unchanged for $R$ ranging from $3$ Mpc to $100$ Mpc.

A comparison of the gravitational masses estimated from
arclike images [eq.(100)] with the masses derived from  hydrostatic
equilibrium [eq.(104)] is shown in Table 6 for four clusters
of galaxies in which both arclike images are detected and the
X-ray data are available (Henry et al., 1982; Gioia and Luppino, 1994):
Abell 370 (A5), MS 1006.0+1202 (arc 4), MS 1008.1-1224 (arc 2) and
MS 1910.5+6736.  It appears that there exists a significant difference between
%**************************Table 6 ****************************************
   \begin{table}
      \caption{Four arc-cluster systems and their masses}
         \label{ }
      \vspace{1cm}
   \end{table}
%**************************************************************************
the virial masses and the gravitational masses from arclike images in all
the four systems.

Three groups have independently announced  similar results for a total of
8 arc-cluster systems (Wu, 1994c; Fahlman et al., 1994; Miralda-Escud\'e and
Babul, 1995), showing that the virial equilibrium has
underestimated the total gravitational masses of clusters of galaxies
by a factor of at least 2.5 up to the arc positions.
There are three main possibilities that may account for the mass discrepancy:
(1)the hot gas in clusters of galaxies may be meanwhile
supported by a non-thermal pressure such as magnetic field (Loeb and Mao,
1994; En$\beta$lin et al., 1995);
(2)cluster matter distributions may be highly prolate with
the long axis along the line of sight (Miralda-Escud\'e and Babul, 1995);
And (3)clusters of galaxies cannot be considered to be the well relaxed
virialized systems (Wu, 1994c).
Nonetheless, the third possibility, if true, may offer
an important  clue to resolving the ``baryon catastrophe" on scale of
clusters of galaxies (White et al., 1993).

The cosmic baryon fraction is the ratio of baryonic
matter $M_b$ (X-ray gas + galaxies) to the total mass $M$
(baryon + non-baryon) of clusters of galaxies: $\Omega_b\equiv M_b/M$,
provided that clusters of galaxies are representative of
the matter distribution of the Universe.
The baryon catastrophe arises because the virial mass $M_v$ derived from
eq.(102) is used as the measurement of the total mass $M$ of cluster of
galaxies, which has led to $\Omega_b$ being $3\sim10$ times larger than
the prediction of the Big Bang Nucleosysthesis and the standard inflation
cosmological model. Now, replacing the virial mass $m_v$ by
the gravitational mass $m_g$ estimated from arclike images
would reduce the baryon fraction $\Omega_b$ by a corresponding factor
of at least  $2.5$ over the region of the typical core radius of cluster.
As a consequence, the ``$\Omega_b$ discrepancy problem"
might vanish.\\

%-------------------------------------------------------------------------
\subsection{Cluster matter distributions from statistical lensing}

Another useful constraint on the matter distribution of galaxy clusters
is provided by
the study of statistical lensing, which attempts to statistically
investigate the possible form of matter distributions of galaxy clusters as a
whole using the properties of giant arcs and/or arclets, such as the total
number, the width and the axial ratio of arcs (Hammer, 1991; Wu and Hammer,
1993; Miralda-Escud\'e, 1993a,b; Grossman and Saha, 1994).  This procedure
is actually a convolution of the magnification probability of galaxy clusters
with the distribution of background galaxies.

The differential probability that a source at $z_s$ is magnified by a factor
within $d{\mu}$ of ${\mu}$ due to a galaxy cluster at $z_d$
is proportional to the lensing cross-section $2\pi D_d^2\beta d\beta$
[eq.(11)]. To find the total magnification
probability $P(z_s,{\mu})$ even for a spherical matter
distribution of cluster of galaxies,
one still needs to know the cluster velocity dispersion
distribution, the length scale distribution and
the spatial distribution.
Unfortunately, all these distributions have not been
very well determined from observations. The X-ray observations of clusters of
galaxies from the Einstein Observatory and EXOSAT
(Edge et al., 1990) as well as
the EMSS (Gioia et al., 1990; Henry  et al., 1992) provide a Schechter
luminosity function at low redshift ($z_d<0.2$) and a power-law
luminosity function $\phi(L_{44})dL_x=KL_{44}^{-\nu}dL_x$
[$L_{44}\equiv (L_x/10^{44}$ ergs s$^{-1}$)] at intermediate redshift
ranging from $0.14$ to $0.6$, showing a significant evolution of
X-ray luminous clusters with cosmic epoch, where $K$ takes different values
at different redshift shells and has units of Mpc$^{-3}$[$L_{44}$]$^{\nu-1}$,
and $\nu$ is the power-law index.  The X-ray luminosity
distribution can be converted into the velocity dispersion distribution
through the correlation between X-ray luminosity $L_x$ and
velocity dispersion $\sigma_v$ (Quintana and Melnick, 1982; Wu and Hammer,
1993):
%106
\begin{equation}
L_x=10^{32.71}\sigma_v^{3.94} {\rm ergs\;s^{-1}}.
\end{equation}
For the distribution of length scale (core radius $r_c$,
effective radius $r_e$, etc.), optical observations
exhibit a constant core radius of $r_c=0.25h_{50}^{-1}$ Mpc for the King model
(Bahcall, 1975;1977) and a constant effective radius of $r_e=6h_{50}^{-1}$
Mpc for the $r^{1/4}$ law (Bears and Tonry, 1986), while X-ray observations
provide a distribution of core radius significantly different from the
constant core (Jones and Forman, 1984):
%107
\begin{equation}
p_c(\log r_c)d\log r_c=\frac{1}{0.32\sqrt{2\pi}}
\exp[-0.5(\log r_c+0.89^2)/0.32^2]\;d\log r_c,
\end{equation}
which leads to an average core radius of $0.17h_{50}^{-1}$ Mpc.  Both the
constant and variable scale length distributions will be adopted in
the computation of $P(z_s, {\mu})$.  As for the number density of
galaxy clusters, both no-evolution and evolution models from the X-ray
data will be used and the respective results will be compared.

The total magnification probability $P(z_s,{\mu})$ for a source at $z_s=2$
by intervening clusters of galaxies with different matter distributions is
shown in Figure 31. Significant differences in $P(z_s,{\mu})$ due to
%@@@@@@@@@@@@@@@@@@@@@@@@@@@@@@@@@@@@@@@@@@@@@@@@@@@@@@@@@@@@@@@@@@@@@@
%                          FIGURE 31
\begin{figure}
 \vspace{1cm}
 \caption{Magnification probability of gravitational lensing
	  by clusters of galaxies for four density profiles.
          The variable $r_c$ distribution of Jones and Forman (1984)
          is adopted for ISC and KING models and a constant $r_e$ of
          $6h_{50}^{-1}$ Mpc is used for the $r^{1/4}$ law. A no-evolution
          scenario is assumed for the number distribution of galaxy clusters.
	  Furthermore, the extension of the background source is neglected.
}
\end{figure}
%@@@@@@@@@@@@@@@@@@@@@@@@@@@@@@@@@@@@@@@@@@@@@@@@@@@@@@@@@@@@@@@@@@@@@@
different models of mass density are clearly seen.
This property opens then a possibility to test
the matter distribution of clusters of galaxies using  statistical lensing.
Convolving $P(z_s,{\mu})$ with the distribution of background galaxies
would yield the number of lensed galaxies
with magnification greater than ${\mu}$.
However, ${\mu}$ measures only the magnification of the apparent luminosity
rather than the geometrical features of the lensed source. The relation
between the magnification ${\mu}$ and the axial ratio $L/W$ of the elongated
image of a background circular galaxy with radius of $R_0$
by a spherical lens can be approximately obtained to be (Wu and Hammer, 1993)
%108
\begin{equation}
\frac{L}{W}={\mu}\left[1+D_0\frac{m(\theta_0)}{\theta_0^2}
          -2\pi D_0\Sigma(\theta_0)\right]^2
\times \frac{\beta_0}{R_0}\sin^{-1}\frac{R_0}{\beta_0},
\end{equation}
in which we have presumed that
$W/\theta_0\ll1$, i.e., the arc width ($W$) is much smaller than the arc
distance ($\theta_0$) from the center of its associated cluster (Apparently,
many detected arcs can meet this condition),
where $\beta_0$ and $\theta_0$ are the alignment parameter and the
corresponding image separation for the center of the background galaxy, and
$D_0=(4G/c^2)(D_dD_{ds}/D_s)$.  In particular, as giant arcs trace
the critical line or the Einstein ring $\theta_E$,
$\theta_0$ can be approximately replaced by $\theta_E$.
Furthermore, one can expand $\sin^{-1}(R_0/\beta_0)$
in $(R_0/\beta_0)$ if the complete ring images are excluded in the statistics
due to their rareness. Then, eq.(108) becomes
%109
\begin{equation}
\frac{L}{W}=4{\mu}(1-\overline{K})^2\left(1-\frac{1}{6}
\frac{R_0^2}{\beta_0^2}+\cdot\cdot\cdot  \right),
\end{equation}
where $\overline{K}=\pi \theta_E^2 D_d^2\Sigma(\theta_E)/m(\theta_E)$.
Numerical computations using the three spherical models in Table 1 show
that the term $(1/6)(R_0/\beta_0)^2$ becomes important only if the radius of
the circular galaxy is larger than $8$ kpc and $\mu>30$, while most of the
known giant arcs have $\mu\sim 10$. Therefore, neglecting the source extension
cannot cause too serious problems in the evaluation of $P(z_s,\mu)$. Thus,
%96
\begin{equation}
{\mu}=\frac{L/W}{4(1-\overline{K})^2}.\\
\end{equation}
This expression relates magnification with axial ratio of
the distorted image and is also very useful to estimate the image magnification
from its geometrical configuration. For example, for the SIS a simple
calculation shows $\overline{K}=1/2$, which then leads to ${\mu}=L/W$.
That is to say, the magnification remains the same as the axial ratio
for the image produced by SIS and thereby, the magnification probability
$P(z_s,{\mu})$ is the probability that a source is elongated by a factor of
greater than $\mu=L/W$.
For other spherical mass density models one can find
the probability $P(z_s,{\mu},L/W)$ from eq.(110) that a source at $z_s$
is magnified
by a factor of ${\mu}$ and  elongated by a factor of $L/W$.

The expected number of giant arcs is finally obtained by convolving the
probability $P(z_s,{\mu},L/W)$ with the Schechter
luminosity function of background galaxies
which is taken from the survey of Broadhurst, Ellis and Shanks (1989). Table 7
gives the result of Wu and Hammer (1993) for the totally expected number
of giant luminous arcs with $B\leq22.5$ and $L/W\geq10$ within $z_s\leq1.25$
over the whole sky. Four mass density models
are used for the matter distribution
of clusters of galaxies which are restricted within the range of
$0.15\leq z_d\leq0.6$.  Both the no-evolution [the Edge et al. (1990) local
X-ray Schechter luminosity function] and the evolution model [the Henry et al.
(1992) X-ray power-law luminosity function]  of galaxy clusters are used.
As a consequence of the significant difference in $P(z_s,{\mu})$ arising
from the various matter distributions (see Figure 31),
the resulting number of giant luminous arcs shows
remarkable differences.
%**************************Table 7 ****************************************
\begin{small}
\begin{table}
	\caption{
The Expected Number of Giant Luminous Arcs over the Whole Sky}
        \label{}
 \[
\begin{array}{|c|c|c|c|c|c|}
\multicolumn{6}{c}{ }\\
\hline
 & & \multicolumn{4}{c|}{\rm number}\\ \cline{3-6}
{\rm model} & {\rm length} &
\multicolumn{2}{c|}{\rm no-evolution} &
                       \multicolumn{2}{c|}{\rm evolution} \\
\cline{3-6}
 & {\rm scale} & \sigma_v\geq800 & \sigma_v\geq1300 &
     \sigma_v\geq800 & \sigma_v\geq1300\\
\hline
  & r_c & 12.5 & 5.9 & 6.5 & 3.7 \\
{\rm ISC} & 0.1r_c  & 123 & 27.5 & 76.6 & 17.1\\
  & 0.05r_c & 152 & 30.0& 92.0& 18.1 \\
\hline
  & r_c & 63.2 & 19.7 & 36.4 & 12.9\\
{\rm KING}& 0.1r_c & 141 & 20.4 & 90.8 & 13.5\\
  & 0.05r_c& 105 & 13.7 & 69.6 & 9.1 \\
\hline
  & r_e & 225 & 77.3 & 123 & 47.6\\
{\rm DEV}& 3h_{50}^{-1}{\rm Mpc}& 462 & 119 & 264 & 73.7\\
  & 1h_{50}^{-1}{\rm Mpc} & 708 & 129 & 432 & 82.1\\
\hline
{\rm SIS} & 0 & 186 & 33.9 & 107 & 18.9\\
\hline
\multicolumn{2}{|c|}{ } &  & & &  \\
\multicolumn{2}{|c|}{\rm clusters of galaxies} & 11000 & 500 & 7100 & 120 \\
\multicolumn{2}{|c|}{ } &  & & &  \\
\hline
\end{array}
  \]
\end{table}
\end{small}
%**************************************************************************
The strong evolutionary scenario of the
X-ray luminous clusters yields fewer clusters than does the no-evolution
model, giving rise to fewer giant luminous arcs predicted from the
evolution model of clusters. Apparently, very rich clusters of
$\sigma_v\geq1300$ km/s have a relatively higher frequency
of producing giant arcs than the rich clusters
of $\sigma_v\geq800$ km/s. A further computation using the clusters of
$\sigma_v\geq1300$ km/s within the redshift interval
$0.15\leq z_d\leq0.4$ gives a frequency of detecting giant luminous arcs
to be as high as $30$--$40\%$. Nevertheless, the model ISC  in Table 7
has underestimated the number of giant luminous arcs by a factor of at least 2
as compared with the known giant luminous arcs
unless the core radii of clusters are reduced significantly, indicative
of a smaller core for the total matter distribution than the X-ray gas.
In the extreme case of the zero-core radius, SIS predicts a number of
giant luminous arcs that marginally reconciles with the observed arcs
in the strong evolution scenario of clusters.  It is then concluded that
the well-known form of ISC, which is found to be the best-fit to the
X-ray luminosity distribution in clusters of galaxies, cannot account
for the giant luminous arcs if dark matter has
the same core radius as the observed one in
X-ray. This implies a more compact dark matter distribution in the central
core of the galaxy cluster.

The statistical study of arc widths has also reached a similar
conclusion (Grossman and Saha, 1994) that cluster mass density profiles
must have core radii at least as small as $\theta_c/\theta_E\leq0.1$ in ISC.
Interestingly, the dynamical analysis of a set of 12 clusters of galaxies
using the X-ray data from the Einstein Imaging Proportional Counter
has found that dark matter appears more peaked in the cluster centers
than the X-ray gas and has core radii of only $50 - 100$ kpc
(Gerbal et al., 1992; Durret et al., 1994).  Although this dynamical
method has given a result that agrees with the lensing estimate, it
is not so clear if the dynamical masses derived from the
assumption of hydrostatic equilibrium for galaxy clusters are
reliable values in view of the argument in section 4.3.

It must be pointed out that the above conclusions drawn from  statistical
lensing by clusters of galaxies should be taken to be very preliminary.
Wu (1993d) has discussed four parameters in the statistical lensing
which may cause some large uncertainties in the predictions of
arc properties due to their limited knowledge from today's observations:
length scale, velocity dispersion, extended source and evolutionary effect.
(1)Core or effective radii:
Both optical and X-ray data are not sufficient to
constrain the distribution of core/effective radii of galaxy clusters.
The optical core radii claimed by Bahcall (1975, 1977) are two times
smaller than those found by Dressler (1978),
whilst the distribution of X-ray core radii eq.(107) shows somewhat larger
scatters.  Alternatively, the constant effective radius of $6h_{50}^{-1}$ Mpc
adopted for the $r^{1/4}$ law is apparently not a good value
for different rich clusters of galaxies.
So, the arc predictions based on the poor data of length scale of
clusters of galaxies  may have large uncertainties. (2)Velocity dispersion:
Magnification probability for lensing is very sensitive to this parameter.
But the average line of sight velocity dispersion along the radius of a
galaxy cluster is a variable rather than a constant except for SIS
(Wu, 1993b; Kochanek, 1993c).  The above statistical lensing takes into
account only the asymptotic value of $\sigma_v$, either close to the center or
at large radius, which may result in a significant variation in lensing
probability. (3)Extended sources: The above point source approximation for
background galaxies would overestimate the magnification
probability $P(z_s,{\mu})$. Recall that a maximum magnification
${\mu}_{max}$ instead of infinity is reached for an extended source (see
section 1.3). Moreover, the consideration of source extensions
might solve the puzzle of too many arclets relative to the giant arcs
predicted from statistical lensing (Bergmann and Petrosian, 1993).
(4)Evolutionary effect:
The strong evolution of galaxy clusters with cosmic epoch would lead to
a number decrease of clusters with redshift, providing much fewer lenses at
high redshift for background galaxies. Conversely, the no-evolution model
is able to produce a relatively larger number of giant luminous arcs in the
clusters of galaxies at high redshift. The observed giant luminous arcs then
open a possibility of testing the evolution model of distant clusters of
galaxies (Wu, 1993a; Bartelmann and Weiss, 1994).
Nevertheless,  the poorly established evolutionary model of clusters
of galaxies may affect the present computation of lensing probability. \\

%-------------------------------------------------------------------------
\subsection{Cluster matter distribution from weak lensing}

The parametric likelihood method of determining the cluster matter
distribution employed in sections 4.3 and 4.4 is suitable for
modeling the giant arcs/arclets, for
constraining the length and shape of global cluster potential wells and
for investigating the relative redshift distribution of the background
objects. The method assumes {\it a priori} a density profile of the cluster
and determines the most likely values of the model parameters using the
properties of the observed arcs/arclets.  It appears that the current
data of the distorted images are still insufficient to discriminate
between the various models. It is, therefore, most desirable that
one can directly derive the cluster matter distribution with a
parameter-free (or non-parametric) method.

Tyson, Valdes and Wenk (1990) firstly detected the coherent distortion
(weak lensing) of faint blue galaxies behind two clusters of galaxies
(A1689 and CL 1409+52).
They used the alignment statistics to extract the lensing signal which is
characterized by an ellipticity (Valdes, Tyson and Jarvis, 1983)
%111
\begin{equation}
e=\frac{a^2-b^2}{a^2+b^2},
\end{equation}
where $a$ and $b$ are the semimajor and semiminor
axes of the principal axis transformed moments. This yields
a positive value of the net alignment for a population of
background galaxies around a foreground cluster, in comparison with
the zero result for a random population of galaxies.
Kaiser and Squires (1993) extended this idea to the outer parts of
clusters and to statistical cluster samples, and developed a
powerful technique to reconstruct the cluster surface mass density
$\Sigma(\mbox{\boldmath $\theta$})$ through the measured distortion $e$,
which is totally independent of the presumed density profile for
clusters. This non-parametric method has been successfully used in
the mapping of two-dimensional mass distributions in several clusters
and is now being further developed for its wide applications in
both weak and strong lensing situations (Seitz and Schneider, 1994;
Kaiser, 1995; Bartelmann, 1995; Bartelmann and Narayan, 1995; etc.).

The image shapes are characterized by the quadrupole moments
(Valdes, Tyson and Jarvis, 1983)
%112,113
\begin{eqnarray}
Q_{ij}=
\frac{\int d^2\theta \theta_i\theta_j I(\mbox{\boldmath $\theta$})}
     {\int d^2\theta I(\mbox{\boldmath $\theta$})}\\
Q^{\prime}_{ij}=
\frac{\int d^2\theta \theta_i\theta_j I^{\prime}(\mbox{\boldmath $\theta$})}
     {\int d^2\theta I^{\prime}(\mbox{\boldmath $\theta$})}
\end{eqnarray}
where $I(\mbox{\boldmath $\theta$})$ and $I^{\prime}(\mbox{\boldmath
$\theta$})$
are the intrinsic and observed surface brightness distributions of a
background source [see also eq.(17)], and the angles are measured from the
centroid of the image.  Assuming that the source is relatively small so
that the magnification matrix $A$ is constant over the images, we have the
following transformation between the quadrupole matrix of source and image
according to eq.(17)
%114
\begin{equation}
Q_{ij}=A_{ik}A_{j\ell}Q^{\prime}_{k\ell},
\end{equation}
or
%115
\begin{equation}
Q=AQ^{\prime}A,
\end{equation}
where $A$ is defined by
$A\equiv \partial \mbox{\boldmath $\beta$}/
\partial \mbox{\boldmath $\theta$}$, and in terms of eqs.(18) and (19)
%116
\begin{equation}
A=\left[
\begin{array}{cc}
1-\kappa-\gamma_1 & -\gamma_2\\
-\gamma_2 & 1-\kappa+\gamma_1
\end{array}  \right].
\end{equation}
By analogy with eq.(111), we define the intrinsic ellipticity
parameter $e$ of the background source and the observed ellipticity
parameter ${e}^{\prime}$ of the image  as
%117,118
\begin{eqnarray}
e=(e_1,e_2)=\left\{\frac{Q_{11}-Q_{22}}{Q_{11}+Q_{22}},\;
                        \frac{2Q_{12}}{Q_{11}+Q_{22}}\right\};\\
e^{\prime}=(e_1^{\prime},e_2^{\prime})
	=\left\{\frac{Q^{\prime}_{11}-Q^{\prime}_{22}}
                             {Q^{\prime}_{11}+Q^{\prime}_{22}},\;
                        \frac{2Q^{\prime}_{12}}
                             {Q^{\prime}_{11}+Q^{\prime}_{22}}\right\},
\end{eqnarray}
which can also be denoted in their complex forms:
$e=e_1+i e_2$  and $e^{\prime}=e_1^{\prime}+i e_2^{\prime}$.

In the limit of weak lensing, one can use the
linear approximation.  Under the transformation eq.(115), eqs.(117)
and (118) reduce to
%119,120
\begin{eqnarray}
e_1^{\prime}=e_1+(\psi_{11}-\psi_{22})(1-e_1^2)-2\psi_{12}e_1e_2;\\
e_2^{\prime}=e_2-(\psi_{11}-\psi_{22})e_1e_2+2\psi_{12}(1-e_2^2).
\end{eqnarray}
The mean intrinsic ellipticity of an ensemble of background galaxies
should be zero: $\langle e_1\rangle=\langle e_2\rangle=
\langle e_1e_2\rangle=0$, while the factors $1-\langle e_1^2\rangle$ and
$1-\langle e_2^2\rangle$ are close to unity in practice. As a result,
the expectation of the image ellipticity is simply
%121,122
\begin{eqnarray}
\langle e_1^{\prime}\rangle = \psi_{11}-\psi_{22};\\
\langle e_2^{\prime}\rangle = 2\psi_{12},\;\;\;\;\;\;
\end{eqnarray}
or
%123
\begin{equation}
\langle e_i^{\prime}\rangle=2\gamma_i,
\end{equation}
i.e., the image distortion is uniquely determined by the tidal
field $\gamma$ of the foreground clusters.  However, the shear
term $\gamma$ is related to the matter term $\kappa$ through
[eqs.(14). (16) and (19)]
%124
\begin{equation}
\gamma(\mbox{\boldmath $\theta$})=
\frac{1}{\pi}\int d^2\mbox{\boldmath $\theta$}^{\prime}
{\chi}(\mbox{\boldmath $\theta$}-
\mbox{\boldmath $\theta$}^{\prime})
\kappa(\mbox{\boldmath $\theta$}^{\prime}),
\end{equation}
where the kernel ${\chi}(\mbox{\boldmath $\theta$})$ is a complex
function (Schneider and Seitz, 1995)
%125
\begin{equation}
{\chi}(\mbox{\boldmath $\theta$})=\frac{\theta_2^2-\theta_1^2
-2i\theta_1\theta_2}{|\mbox{\boldmath $\theta$}|^4}.
\end{equation}
Finally, the inversion of eq.(124) yields
the cluster surface mass density
$\Sigma(\mbox{\boldmath $\theta$})$,
which is expressed by the observed ellipticity parameter of the images,
%126
\begin{equation}
\Sigma(\mbox{\boldmath $\theta$})=\frac{\Sigma_c}{\pi}
 \int d^2 \mbox{\boldmath $\theta$}^{\prime}
      {\it Re}[{\chi}^*(\mbox{\boldmath $\theta$}-
             \mbox{\boldmath $\theta$}^{\prime})
	     e^{\prime}(\mbox{\boldmath $\theta$}^{\prime})],
\end{equation}
where $\chi^*$ is the complex conjugation of $\chi$ and
${\it Re}$ takes the real part of the complex variable.

Since the first detection of gravitational weak shear out to a radius
of 3.0 $h_{50}^{-1}$ Mpc in CL 0024+1654 (Bonnet, Mellier and Fort,
1994), several measurements of the weak distortion of background galaxies
have been made and  the applications of the lensing inversion technique
to the observed data have turned to be very successful in the reconstruction
of the mass distribution of clusters (Fahlman et al., 1994;
Smail et al. 1995; Smail and Dickinson, 1995; Tyson and Fischer, 1995;
Squires et al., 1995; Kneib et al., 1995; etc.).
Although there are some disagreements
about the resulting gravitating masses
of clusters, for instance, Fahlman et al. (1994) derived a gravitational
mass of 2.5 -- 3 times larger than its virial mass in cluster MS 1224 while
Squires et al. (1995) found accordance between the two masses in Abell
2218, it appears very promising that one can precisely determine the cluster
mass from lensing method
in the near future by improving the inversion
technique (Bartelmann, 1995; Schneider, 1995; Schneider and Seitz, 1995;
Seitz and Schneider, 1995a; Kaiser, 1995).\\

%---------------------------------------------------------------------
\subsection{Quasar-cluster associations}

The presence of giant arcs and arclets associated with gravitational
potentials of clusters of galaxies indicates that clusters of
galaxies are very efficient lenses. Actually, it was noticed
at the same time  when giant luminous arcs were discovered that
some of the high-redshift 3CR galaxies
are gravitationally magnified by the low-redshift clusters of
galaxies lying in their lines of sight (Hammer, Nottale and Le F\`evre,
1986; Le F\`evre, Hammer and Jones,
1988; Le F\`evre et al., 1988; Hammer and Le F\`evre, 1990).
Motivated by these observations, Wu and Hammer (1993) even
explored the possibility of whether there are radio arcs behind
galaxy clusters.

What would happen to background quasars if foreground clusters of galaxies
act as lenses ?  Four recent measurements have answered this question
by discovering a significant quasar overdensity behind
foreground clusters using different quasar and cluster samples:
(1)the Large Bright Quasar Survey and Zwicky
clusters (Rodrigues-Williams and Hogan, 1994), (2)the 1 and 2 Jy Radio
Source Surveys and Abell clusters (Wu and Han, 1995),
(3)the variability selected quasars and clusters (Rodrigues-William and
Hawkins, 1995). (4)the 1 Jy Radio Source Catalog and Zwicky clusters
(Seitz and Schneider, 1995b). Table 8 summarizes these searches and
their resulted quasar overdensity density  $q$, in which only
the most significant $q$ measured at a fixed position $\theta$ and
a limiting magnitude (or flux) is given. Figure 32 shows
the variations of $q$ for Abell clusters at $z_d<0.2$ around
2 Jy radio quasars at $z_s>0.5$.
%**************************Table 8 ****************************************
   \begin{table}
      \caption{Quasar-cluster associations: observations and models}
	\label{}
      \vspace{1cm}
   \end{table}
%**************************************************************************
%@@@@@@@@@@@@@@@@@@@@@@@@@@@@@@@@@@@@@@@@@@@@@@@@@@@@@@@@@@@@@@@@@@@@@@
%                          FIGURE 32
\begin{figure}
 \vspace{1cm}
 \caption{Variations of enhancement factor of Abell clusters around
          2 Jy radio quasars. The data have been normalized at $3^o$
          and the standard deviation ($1\sigma$) is plotted in
          each measurement.
          The fit to the prediction by a singular isothermal sphere
          as lensing object is shown using a critical radius of
          $0.2^o$.}
\end{figure}
%@@@@@@@@@@@@@@@@@@@@@@@@@@@@@@@@@@@@@@@@@@@@@@@@@@@@@@@@@@@@@@@@@@@@@@

In a similar way to the study of  quasar-galaxy associations,
we can evaluate  the enhancement factor $q$ for the quasar-cluster
associations. Given a magnification
$\mu$, a general expression  is available for both
optically- and radio-selected quasars
%127
\begin{equation}
q=\frac{N(<m+2.5\log {\mu})}{N(<m)}\;\frac{1}{\mu}\;=\;
\frac{N(>S/{\mu})}{N(>S)}\;\frac{1}{\mu}
\end{equation}
where $m$ and $S$ denote the limiting magnitude and the flux threshold,
respectively.
The surface number density of optically-selected quasars has been
given by BSP [see eq.(94)]. For the radio source counts $N(>S)$,
a least-square fit of a power-law to the observations at
5 GHz (Langston et al., 1990; Fomalont et al., 1991) yields
%129
\begin{equation}
N(>S)=\left\{
\begin{array}{ll}
1.27\times 10^{6}\;S^{-1.46}, & S>10\;{\rm mJy};\\
2.10\times10^5\;S^{-1.10}, & S<10\;{\rm mJy},
\end{array} \right.
\end{equation}
where $S$ is the units of mJy. Note that radio source catalogs are
composed not only of galaxies but also quasars. The fraction of quasars
in radio source surveys varies with flux threshold. Therefore,
the employment of $N(>S)$ in the study of quasar-cluster associations
provides only an estimate of $q$.

Now we work with the lensing models of clusters of galaxies and/or
their associated matter inhomogeneities and test whether one
can explain the reported quasar overdensity behind clusters
on scale of $\sim10$ arcminutes in terms of gravitational lensing.
We take an average enhancement $\langle q\rangle$ [eq.(93)]
over the search range of $\theta$
around clusters instead of the local enhancement $q$ at $\theta$.
We further assume a flat cosmological model $\Omega=1$ and adopt
$H_0=50$ km/s/Mpc.

Conventionally, clusters of galaxies should be considered to be the
lensing objects  for the reported quasar-cluster associations.
Utilizing SIS as the mass model and its magnification of
eq.(27), we can estimate the cluster velocity dispersion $\sigma_c$
that is required to produce the observed $\langle q\rangle$.
Figure 32 plots such a fit to the observed data. Surprisingly,
the best fitted Einstein radius in this example is $\sim0.2^o$,
corresponding to a velocity dispersion of $\sigma_c\approx5000$ km/s
if the typical redshifts of Abell clusters and of radio sources
are taken to be 0.1 and 1, respectively.  The similar results
are found for the rest three measurements (Table 8).
Apparently, the masses ($\sim\sigma_c^2$)
that are needed to produce the four measured
enhancement factors are substantially larger than the
realistic value for clusters.

It has been known for some years that the weak lensing by large-scale matter
inhomogeneities may contribute a significant effect on the
background quasars. It may be the cause for the
quasar-galaxy associations observed on the similar
scale ($\sim10^{\prime}$) (Fugmann, 1988;1990; Bartelmann and Schneider,
1993a,b;1994). We can now work out how large an additional mass surface
density from the large-scale matter clumps that clusters of galaxies trace
is need to produce the quasar-cluster associations.
To do this, we add a uniform mass sheet $\Sigma$ to clusters of galaxies.
It turns out that the Einstein radius $\theta_E$ and the image
separation are increased by a factor of $(1-\Sigma/\Sigma_{c})^{-1}$
and the lensing magnification becomes
%129
\begin{equation}
\mu=\frac{1}{|1-\theta_E/\theta|(1-\Sigma/\Sigma_{c})^2},
\end{equation}
where $\Sigma_c$ is
the critical surface mass density that any lens must exceed in order
to produce multiple images by itself [eq.(15)].
Quantitatively, the minimum $\Sigma_{c}$ for a source at $z_s=2$
is $0.41$ g/cm$^2$.  In Table 8 we give the required surface mass
density $\Sigma$ for each of the measurements.  Note that $\Sigma$
deduced from the radio selected
quasars is a factor of $\sim2$ larger than the one from
the optically selected samples. This is due to the contamination of
radio galaxies in the radio source catalog. The result from Seitz and
Schneider (1995) illustrates very well this effect.

We can estimate the matter contribution from all the galaxy clusters
that follow the cluster spatial two-point correlation function
$\xi(r/r_{cc})^{-1.8}$, where the correlation amplitude is
$r_{cc}=40$ Mpc (Postman, Huchra and Geller, 1992). If we assume that
each cluster has a SIS mass density profile and a gravitational
radius $R_c$ and the mean number density of clusters is constant,
the surface mass density of clusters enclosed within
$\theta$ around a given cluster at $z_d$ is (Wu and Fang, 1995)
%130
\begin{equation}
\Sigma(\theta)=4n_0M_cr_{cc}(1+z_d)^3\;F(\theta,r_{cc},R_c),
\end{equation}
where $n_0$ is the present cluster number density,
$M_c=2\sigma_c^2R_c/G$ is the total cluster mass and $F$ is a function
given by $\xi(r)$ and SIS model. Numerical computations show
that $F\approx2\sim3$ for $R_c=3$--$5$ Mpc over the range of
$\theta=1^{\prime}$--$80^{\prime}$. Let $\Omega_c$ denote the
fraction of the total cluster matter in the mass density of
the Universe, we have
%131
\begin{equation}
\Sigma=0.01\Omega_c\left(\frac{(1+z_d)^3}{1.15}\right)
           \left(\frac{F}{3}\right)\;{\rm g/cm}^2.
\end{equation}
Unfortunately, this surface mass density provided by all
galaxy clusters following $\xi(r)$ is  an order of magnitude lower
than that required for the
quasar-cluster associations even if $\Omega_c=1$.

The mass surface density from large-scale structures of the Universe
can be estimated through
%132
\begin{equation}
\Sigma=\int [\rho(r)-\rho_0]dr\sim1.45\times10^{-3}\delta
            \left(\frac{r}{100\;{\rm Mpc}}\right)
	    \; {\rm g/cm}^2,
\end{equation}
in which $\delta$ is the mean density contrast over scale of $R$.
However, the evaluation of $\Sigma$ is sharply constrained
by the measurements of temperature anisotropy $\Delta T/T$
of the cosmic
background radiation on various scales. Numerical computation
indicates that it is impossible to attribute the large mass
surface density derived from the quasar-cluster associations to
any matter clumps on scale of $R>20$ Mpc in the Universe
if $\Delta T/T=1\sim5\times10^{-5}$ (Wu and Fang, 1995).

So, if the reported associations between background bright quasars
and foreground clusters of galaxies are not due to statistical
variations arising from the quasar/cluster selections and patchy
Galactic obscuration, we need to consider the following possibilities:
(1)There may exist a large amount of unseen matter between clusters
of galaxies on scale of $\sim10$ Mpc, because the above calculations
did not include the unbound cluster matter. This can be
tested using the N-body simulations, as was made by Bartelmann and
Schneider (1993b) for quasar-galaxy associations. (2)The working
hypothesis may be wrong, i.e., the observed background quasar counts
may deviate from their intrinsic ones.  On the scale
of galaxies, Schneider (1992) has demonstrated that dropping the
unaffected background hypothesis does not significantly
improve the situation in quasar-galaxy associations. Whether
the cluster matter or large-scale structures would contribute
a non-negligible effect on the quasar number counts needs to
be further investigated.\\

\bigskip
%@@@@@@@@@@@@@@@@@@@@@@@@@@@@@@@@@@@@@@@@@@@@@@@@@@@@@@@@@@@@@@@@@@@@@@@
\bigskip
\begin{large}
{\noindent}{\bf FINAL REMARKS}\\
\end{large}

{\noindent}The past few years have been exciting times for lensing people. In
particular, the microlensing experiments have detected a few ten events
associated with the compact objects of the Galactic halo/disk and/or the
LMC halo/disk, indicative of the success of using gravitational lensing
effect for the searches of dark matter, which will have a strong impact on
various aspects of cosmology study today.  The rapidly increasing number
of new lens systems (multiple quasars/galaxies, radio rings/galaxies,
arcs/clusters of galaxies, etc.) has made it possible to
study the matter distribution of the Universe statistically
and to determine the
cosmological parameters ($H_0$, $\Omega_0$ and $\lambda_0$). This
is of particularly significance since lensing provides not only an independent
means to evaluate these important issues in cosmology but also a test for
the validity of other astronomical/physical methods. The determination
of the Hubble constant $H_0$ from the time delay of  double quasars
and mapping matter distribution in clusters of galaxies with luminous
arcs and arclets are the two excellent examples of the lensing applications
in cosmology.

Indeed, the study of gravitational lensing has been developed so rapidly
in both theory and observation, and it is not practical to summarize
every subject in this review destined for the readers who are not the experts
in lensing.  In the present article it is even impossible to include some
new discoveries, new observations and new theories which appeared during
the writing of the article. The interested readers are recommended to
refer to the recent reviews on the lensing applications in cosmology
(Blandford and Narayan, 1992; Schneider, 1996),
on the lensing observations (Refsdal and
Surdej, 1994), on the arc(let)s in clusters of galaxies (Fort and Mellier,
1994), in particular, the excellent monograph of gravitational lensing
by Schneider, Ehlers and Falco (1992).

Two quotes can be used as the final remarks of this review on gravitational
lensing:\\

\begin{it}
``An astronomer can use beams of photons to probe a condensation of
{\bf dark matter} in much the same way that a nuclear physicist used
beams of electrons to study the structure of an atomic nucleus. "
\end{it}
(Blandford and Kochanek, 1987)\\

\begin{it}
``A galactic gravitational lens can be used as the ultimate astronomical
telescope."
\end{it}
(McBreen and Metcalfe, 1987)\\

\bigskip
%@@@@@@@@@@@@@@@@@@@@@@@@@@@@@@@@@@@@@@@@@@@@@@@@@@@@@@@@@@@@@@@@@@@@@@@
\bigskip

{\noindent}ACKNOWLEDGEMENTS \\

Most of this research was made while I was taking my postdoctoral
position at DAEC of Observatoire de Paris-Meudon,
during the period June 1990 -- January 1994.
I first want to thank all of my friends at Meudon for their
friendship and assistance, namely,
Catarina, Chantal, Claudia, Christain, Delphine, Dominique, Francois, Gary,
Guida, Ilidio, Josefine, Jose-Luis,  Laurence,
Isabel, Lourdes, Maarten,  Maria, Marie-Christine, Veerle.
I wish to acknowledge Francois Hammer,  Ren\'e Racine, Hanns Ruder,
Genevieve Soucail, Rolf Stabell,  Joachim Wambsganss, and  Corvin Zahn
for providing photographs and significant help.
Jiansheng Chen, Zugan Deng, Li-Zhi Fang, Qibin Li
and Zhenlong Zou are appreciated for
their encouragement and continued supports for my research at Beijing
Astronomical Observatory and in the University of Arizona.
I would like to give my special appreciation to
Peter Schneider for careful reading the manuscript,
for pointing out many mistakes and for numerous and valuable
suggestions and comments. Finally, I am grateful to my dear friend Jie
for her companionship in the final stage of this work.
This work was supported by the National Science Foundation of China.\\

\bigskip
%@@@@@@@@@@@@@@@@@@@@@@@@@@@@@@@@@@@@@@@@@@@@@@@@@@@@@@@@@@@@@@@@@@@@@@@

\newpage

%####################### Table 1 ######################################
\begin{tabular}{|c|c|c|c|}
\multicolumn{4}{c}{Table 1 ~ Gravitational Lensing Models}\\
\multicolumn{4}{c}{ }\\
\hline
   &   &   &  \\
model & ISC & KING & $r^{1/4}$ law\\
   &   &   &  \\
\hline
   &   &   &  \\
surface density& $\frac{\Sigma_0}{\sqrt{1+\theta_0^2}}$
               & $\frac{\Sigma_0}{1+\theta_0^2}$
               & $\Sigma_0\exp[-7.669\theta_0^{1/4}]$\\
   &   &   &  \\
\hline
   &   &   &  \\
length scale & $r_c$ & $r_c$ & $r_e$\\
   &   &   &  \\
\hline
central density ($\rho_0$)  &   &  &  \\
or total mass ($M_0$) \&  & $\rho_0=\frac{\sigma_{v}^2}{2\pi Gr_c^2}$
                          & $\rho_0=\frac{9\sigma_{v}^2}{4\pi G r_c^2}$
                          & $M_0=9.0\frac{r_e\sigma_{v}^2}{G}$\\
velocity dispersion ($\sigma_{v}$)  &  &  &  \\
\hline
   &   &   &  \\
lensing equation & $\beta_0=\theta_0-D\frac{\sqrt{1+\theta_0^2}-1}{\theta_0}$
                 & $\beta_0=\theta_0-D\frac{\ln(1+\theta_0^2)}{\theta_0}$
                 & $\beta_0=\theta_0-D\frac{m_0(\theta_0)}{\theta_0}$ ~ (*)\\
   &   &   &  \\
\hline
   &   &   &  \\
$D$ & $4\pi\frac{\sigma_{v}^2}{c^2}\frac{D_dD_{ds}}{r_cD_s}$
    & $18\frac{\sigma_{v}^2}{c^2}\frac{D_dD_{ds}}{r_cD_s}$
    & $36\frac{\sigma_{v}^2}{c^2}\frac{D_dD_{ds}}{r_eD_s}$\\
   &   &   &  \\
\hline
   &   &   &  \\
critical $D$ & 2 & 1 & $3.370\times10^{-3}$\\
   &   &   &  \\
\hline
\multicolumn{4}{l}{  }\\
\multicolumn{4}{l}{  }\\
\multicolumn{4}{l}{
$
^* \;\;m_0(\theta_0)=1-\exp\left(-7.669\theta_0^{1/4}\right)
\displaystyle\sum_{n=0}^{7}\frac{1}{n!}
\left(7.669\theta_0^{1/4}\right)^n
$
}\\
\multicolumn{4}{l}{  }\\
\end{tabular}

\newpage

%####################### Table 3 ######################################
\begin{tabular}{lcccccllc}
\multicolumn{9}{c}{Table 3 ~Gravitationally-Lensed Multiple Quasars$^*$}\\
\multicolumn{9}{c}{ }\\
\hline
\multicolumn{9}{c}{ }\\
\multicolumn{1}{c}{name} & image No. & lens & $z_d$ & $z_s$ &
$\Delta\theta_{max}$ & \multicolumn{1}{c}{status} &
\multicolumn{1}{l}{discovers} & year\\
\multicolumn{9}{c}{ }\\
\hline
\multicolumn{9}{c}{ }\\
0957+561 & 2 & G+C & 0.36,0.5 & 1.41 & $6''.1$ & confirmed  &
                                                   Walsh et al. & 1979\\
\multicolumn{9}{c}{ }\\
1115+080 & $\geq4$ & G & 0.29 & 1.72 & $2''.3$ & confirmed   &
                                                   Weymann et al. & 1980\\
\multicolumn{9}{c}{ }\\
2345+007 & 2 & G(?) & 1.49(?) & 2.15 & $7''.3$ & possible
&Weedman et al.& 1982\\
\multicolumn{9}{c}{ }\\
1634+267 & 2 & G(?) & 0.57(?) & 1.96 & $3''.8$ & possible &
Djorgovski \& Spinrad & 1984\\
\multicolumn{9}{c}{ }\\
2016+112 & 3 &  G & 1.01(?) & 3.27 & $3''.8$ & confirmed   &
                                  Lawrence et al.  & 1984\\
\multicolumn{9}{c}{ }\\
2237+0305 & 4 & G & 0.04 & 1.69 & $1''.8$ & confirmed   &
                                             Huchra et al. &1985\\
\multicolumn{9}{c}{ }\\
0142-100 & 2 & G & 0.49 & 2.72 & $2''.2$ & confirmed   &
                                             Surdej et al.  &1987\\
\multicolumn{9}{c}{ }\\
1413+117 & 4 & G(?) & 1.4(?) & 2.55 & $1''.1$ & confirmed   &
Magain et al.  &1988\\
\multicolumn{9}{c}{ }\\
1120+019 & 2 & G(?)C(?)& 0.6(?) & 1.46 & $6''.5$ & possible &
Meylan \& Djorgovski & 1989\\
\multicolumn{9}{c}{ }\\
1429-008 & 2 & (?) & 1.6(?) & 2.08 & $5''.1$ & possible &
Hewett et al.  &1989\\
\multicolumn{9}{c}{ }\\
0952-0115 & 2& (?) & (?) & 4.5 & $0''.9$ & possible &
McMahon et al. & 1992\\
\multicolumn{9}{c}{ }\\
1208+1011 & 2 & (?) & (?) & 3.80 & $0''.47$ & possible &
                                          Magain et al. &1992\\
\multicolumn{9}{c}{ }\\
1422+231 & 4 & G(?) & 0.64(?) & 3.62 & $1''.3$ &
confirmed & Patnaik et al.&1992\\
\multicolumn{9}{c}{ }\\
1009-025 & 2 & G(?) & 1.62(?) & 2.74& $1''.55$ & possible &
                          Surdej et al. &1993\\
\multicolumn{9}{c}{ }\\
1104-1805 & 2 & (?) & 1.66(?) & 2.30 & $3''$ & possible &
                                                    Wisotzki et al. &1993\\
\multicolumn{9}{c}{ }\\
0240-343 & 2  &  G(?) & 0.34(?) & 1.4 & $6''.1$ &
possible & Tinney & 1995\\
\multicolumn{9}{c}{ }\\
J03.13 & 2 & G(?) & 1.09(?) & 2.55 & $0''.84$ & possible &
Claeskens et al. & 1995\\
\multicolumn{9}{c}{ }\\
%1608+656 & 4 & G & 0.630 & (?) & $2''.1$ & possible &
%                                                    Myers et al. &1995\\
%\multicolumn{9}{c}{ }\\
%1600+434 & 2 & G(?) & ? & 1.61(?) & 1''.4 & possible &
%                                                    Jackson et al. &1995\\
\multicolumn{9}{c}{ }\\
\hline
\end{tabular}

\bigskip

{\noindent}  $^*$ Radio sources are not included.
{}~~G=galaxy; ~~C=cluster; ~~$\Delta\theta$=separation

\newpage

%####################### Table 5 ######################################

\begin{tabular}{lllclll}
\multicolumn{7}{c}{\Large Table 5 ~~Arclike Images}\\
\multicolumn{7}{c}{ }\\
\hline
\multicolumn{7}{c}{ }\\
\multicolumn{1}{l}{arc cluster} & $z_d$ & $\sigma_v$(km/s) &
$L_{x,44}({\rm ergs/s})$ & arc redshift($z_s$) & discovers & year \\
\multicolumn{7}{c}{ }\\
\hline
\multicolumn{7}{c}{ }\\
Abell 222   & 0.213 & 570  & 3.7   &            & Smail et al. & 1991 \\
Abell 370   & 0.374 & 1364 & 9.7   & 0.725,1.3? & Soucail et al. & 1987a \\
Abell 963   & 0.206 &      & 9.1   & 0.771      & Lavery \& Henry & 1989\\
Abell 1689  & 0.196 & 1989 & 17.   &            & Tyson et al. & 1990 \\
Abell 1942  & 0.224 &      &       &            & Smail et al. & 1991\\
Abell 2104  & 0.155 &      & 8.0   &            & Pierre et al. & 1994\\
Abell 2163  & 0.203 &      &       &0.728,0.742 & Soucail et al. & 1994\\
Abell 2218  & 0.176 &     & 6.5   &0.702,1.034 & Pello-Descayre et al. & 1988\\
Abell 2219  & 0.225 &      & 18.   & $\sim1$    & Smail et al. & 1995\\
Abell 2280  & 0.326 &      & 5.1   &            & Gioia et al. & 1995\\
Abell 2390  & 0.231 &     &       & 0.913      & Pello-Descayre et al. & 1991\\
Abell 2397  & 0.212 &      &       &            & Smail et al. & 1991\\
Abell S295  & 0.301 & 900  &       &            & Edge et al. & 1994\\
\multicolumn{7}{c}{ }\\
\hline
\multicolumn{7}{c}{ }\\
%Cl 0018-20  & 0.27  & 1300 &       &            &       & 1993 \\
Cl 0024+1654& 0.391 & 1300 & 2.7   &  1.39?     & Koo & 1987\\
Cl 0302+1658& 0.426 &      & 5.0   &            & Mathez et al. & 1992\\
Cl 0500-24  & 0.316 & 1375 &       &  0.91?     & Giraud & 1988\\
Cl 1409+52  & 0.46  & 3000 & 9.2   &            & Tyson et al. & 1990\\
Cl 2236-04  & 0.56  &      &       &  1.116     & Melnick et al. & 1993\\
Cl 2244-02  & 0.336 &      & 1.5   &  2.237     & Soucail et al. & 1987a\\
\multicolumn{7}{c}{ }\\
\hline
\multicolumn{7}{c}{ }\\
MS 0440+0204& 0.190 &      & 4.0   &            & Luppino et al. & 1993\\
MS 0451-0305& 0.55  &      & 20.   &            & Le F\`evre et al. & 1994\\
MS 1006+1202& 0.221 &      & 4.8   &            & Le F\`evre et al. & 1994\\
MS 1008-1224& 0.301 &      & 4.5   &            & Le F\`evre et al. & 1994\\
MS 1455+2232& 0.259 &      & 16.   &            & Le F\`evre et al. & 1994\\
MS 1621+2640& 0.426 &      & 4.5   &            & Le F\`evre et al. & 1994\\
MS 1910+6736& 0.246 &      & 4.4   &            & Le F\`evre et al. & 1994\\
MS 2053-0449& 0.583 &      & 5.8   &            & Le F\`evre et al. & 1994\\
MS 2137-2353& 0.313 &      & 16.   &            & Fort et al. & 1992\\
MS 2318-2328& 0.187 &      & 6.8   &            & Le F\`evre et al. & 1994\\
\multicolumn{7}{c}{ }\\
\hline
\multicolumn{7}{c}{ }\\
AC 114      & 0.31  & 1649 & 4.0   &    0.639   & Smail et al. & 1991\\
0956+561$^*$   &0.36,0.5&      &       &            & Bernstein et al. & 1993\\
GHO 2154+0508& 0.32  &      &       &   0.721    & Lavery et al. & 1993\\
PKS0745-191 & 0.1028 &    &   &         0.433   & Allen et al. & 1995\\
RXJ1347.5-1145 & 0.451 &      & 62. &     &       Schindler et al. & 1995\\
ACO 3408 &       0.042 &      &     & 0.073  & Campusano \& Hardy & 1995\\
\multicolumn{7}{c}{ }\\
\hline
\end{tabular}
\begin{small}
{\noindent}$^*$ The recent observation (Dahle, Maddox and Lilje,
1994) suggests that this ``arc system" is the result of chance aligments of
three and two different objects, and not gravitationally lensed arcs.
\end{small}

\newpage

%####################### Table 6 ######################################

\begin{tabular}{|c|c|c|c|c|c|c|c|c|c|}
\multicolumn{10}{c}
{\Large\bf Table 6 ~~Four arc-cluster systems and their masses}\\
\multicolumn{10}{c}{ }\\
\multicolumn{10}{c}{ }\\
\multicolumn{10}{c}{ }\\
\hline
\multicolumn{3}{|c|}{ } & \multicolumn{4}{|c|}{ } & \multicolumn{3}{|c|}{ }\\
\multicolumn{3}{|c}{\large cluster} & \multicolumn{4}{|c|}{\large arc} &
\multicolumn{3}{c|}{\large mass}\\
\multicolumn{3}{|c|}{ } & \multicolumn{4}{|c|}{ } & \multicolumn{3}{|c|}{ }\\
\hline
          &       &     &    &   &          &      &      &      &     \\
name & $z_d$ & $L_{x,44}$ &
$\theta(^{\prime\prime})$ & $L(^{\prime\prime})$ & $L/W$ & $z_s$ &
 $m_v(\theta)\;(M_{\odot}) $
 & $m_g(\theta) \;(M_{\odot})$  & $m_g(\theta)/m_v(\theta)$\\
          &       &     &    &   &          &      &      &      &     \\
\hline
          &       &     &    &   &          &      &      &      &     \\
Abell 370 & 0.374 & 9.7 & 56 & 9 & 18 & 1.3(?)
& $2.26\cdot10^{14}$ & $8.20\cdot10^{14} $ & 3.63\\
          &       &     &    &   &          &      &      &      &     \\
\hline
          &       &     &    &   &          &      &      &      &     \\
          &       &     &    &   &  & 0.6
&  $1.36\cdot10^{14} $ & $6.91\cdot10^{14} $ & 5.08\\
MS1006.0+1202&0.221 & 4.819 & 62 &  4.9&7.0           &      &      &      &
 \\
          &       &     &    &   & &  2.0
& $1.36\cdot10^{14} $ & $4.87\cdot10^{14} $ & 3.58\\
          &       &     &    &   &          &      &      &      &     \\
\hline
          &       &     &    &   &          &      &      &      &     \\
          &       &     &    &   & & 0.6
& $1.33\cdot10^{14} $ & $7.47\cdot10^{14} $ & 5.60\\
MS1008.1-1224&0.301& 4.493 & 51  &  4.0&6.5           &      &      &      &
 \\
          &       &     &    &   & & 2.0
& $1.33\cdot10^{14} $ & $4.34\cdot10^{14} $ & 3.25\\
          &       &     &    &   &          &      &      &      &     \\
\hline
          &       &     &    &   &          &      &      &      &     \\
          &       &     &    &   &  & 0.6
& $1.63\cdot10^{14} $ & $9.95\cdot10^{14} $ & 6.12\\
MS1910.5+6737&0.246 & 4.386 & 67 &  6.1&10.5          &      &      &      &
 \\
          &       &     &    &   &  & 2.0
&  $1.63\cdot10^{14} $ & $6.64\cdot10^{14} $ & 4.08\\
          &       &     &    &   &          &      &      &      &     \\
\hline
\end{tabular}

\newpage

%####################### Table 8 ######################################

\begin{tabular}{cccccccccc}
\multicolumn{10}{c}{{\large Table 8 ~~Quasar-Cluster Associations:
			Observations and Models}}\\
\multicolumn{10}{c}{ }\\
\multicolumn{10}{c}{ }\\
\hline
\hline
\multicolumn{10}{c}{ }\\
clusters & quasars   & $\langle z_d\rangle^{a}$
 & $\langle z_s\rangle^{b}$ & $\theta^{c}$  &
$\langle q\rangle_{obs}$ & $(\sigma_c/10^3)^d$ &
$(\sigma_c/10^3)^2$ & $\Sigma^{e}$ & ref$^f$\\
\multicolumn{10}{c}{ }\\
\hline
\multicolumn{10}{c}{ }\\
Zwicky & $B\leq18.5$ & 0.2 & 1.8 & 52 & $1.7_{-0.4}^{+0.5}$ &
         $5.3_{-1.6}^{+1.6}$ & $28^{+20}_{-14}$  &
	 $0.10_{-0.05}^{+0.04}$  & 1 \\
\multicolumn{10}{c}{ }\\
Abell  & $S\geq2$ Jy & 0.1 & 2.0 & 24 & $1.7_{-0.5}^{+0.5}$ &
         $4.7_{-1.8}^{+1.2}$ & $22^{+13}_{-14}$ &
	 $0.28_{-0.18}^{+0.10}$  & 2 \\
\multicolumn{10}{c}{ }\\
UKJ287$^{g}$ & $B\leq18.5$ & 0.15 & 1.5 & 7.2 & $2.0_{-0.2}^{+0.2}$ &
         $2.3_{-0.2}^{+0.2}$ &  $5.3^{+1.0}_{-0.9}$ &
	 $0.12_{-0.02}^{+0.02}$  & 3 \\
\multicolumn{10}{c}{ }\\
Zwicky & $B\leq19$ & 0.2 & 1 & $78$ & $\sim1.3$ &
         $4.3$ & 18 & $0.06$  & 4 \\
       & $S\geq1$ Jy &  &  &  &  &
         $5.6$ & 31 & $0.11$   &  \\
\multicolumn{10}{c}{ }\\
\hline
\end{tabular}

\bigskip
\bigskip

$^a$Mean cluster redshift

$^b$Mean quasar redshift

$^c$Search range in arcminutes

$^d$Required cluster velocity dispersion in
	          units of 1000 km/s

$^e$Required surface mass density in g/cm$^2$ for
	         $\Omega=1$ and $H_0=50$ km/s/Mpc.

$^f$References -- (1)Rodrigues-Williams and Hogan, 1994;
		(2) Wu and Han, 1995; (3) Rodrigues and Hawkins, 1995;
                (4) Seitz and Schneider, 1995.

$^g$Clusters in UKJ287 field

\end{document}